\documentclass[iop]{emulateapj}

\usepackage{color}
\usepackage{amssymb, amsmath}

\usepackage[utf8]{inputenc}
\usepackage{lipsum}
\usepackage{pifont}
\usepackage{amsmath, amssymb, bm}
\usepackage{balance}
\usepackage{multirow}
\usepackage{mathtools}
\usepackage{enumitem}
\usepackage{booktabs}
\usepackage{appendix}
\usepackage{slashed}
\shorttitle{Detecting neutrino mass by combining matter clustering, halos, and voids}
\shortauthors{Bayer, Villaescusa-Navarro et al.}

\usepackage[usenames,dvipsnames]{xcolor}
\usepackage{hyperref}
\hypersetup{
    colorlinks = true,
    citecolor = {MidnightBlue},
    linkcolor = {BrickRed},
    urlcolor = {blue}		
}

\newcommand{\be}{\begin{equation}}
\newcommand{\ee}{\end{equation}}
\newcommand{\ba}{\begin{eqnarray}}
\newcommand{\ea}{\end{eqnarray}}

\begin{document}

\title{Detecting neutrino mass by combining matter clustering, halos, and voids}

\author{
Adrian E.~Bayer$^{1,2,*}$, Francisco Villaescusa-Navarro$^{3,4,\dagger}$, Elena Massara$^{5,4}$, Jia Liu$^{1,2,6}$, David N.~Spergel$^{3,4}$, Licia Verde$^{7,8}$, Benjamin D.~Wandelt$^{9,10,4}$, Matteo Viel$^{11,12,13,14}$, Shirley Ho$^{4,3,15}$\\}
\affil{$^1$ Berkeley Center for Cosmological Physics, University of California, 341 Campbell Hall, Berkeley,  CA 94720, USA}
\affil{$^2$ Department of Physics, University of California, 366 LeConte Hall, Berkeley,  CA 94720, USA}
\affil{$^3$ Department of Astrophysical Sciences, Princeton University, Peyton Hall, Princeton NJ 08544-0010, USA}
\affil{$^4$ Center for Computational Astrophysics, Flatiron Institute, 162 5th Avenue, 10010, New York, NY, USA}
\affil{$^5$ Waterloo Centre for Astrophysics, University of Waterloo, 200 University Ave W, Waterloo, ON N2L 3G1, Canada}
\affil{$^6$ Lawrence Berkeley National Laboratory, 1 Cyclotron Road, Berkeley, CA 94720, USA}
\affil{$^7$ Institut de Ci\`{e}ncies del Cosmos, University of Barcelona, ICCUB, Barcelona 08028, Spain}
\affil{$^8$ Institucio Catalana de Recerca i Estudis Avancats, Passeig Lluis Companys 23, Barcelona 08010, Spain}
\affil{$^9$ Institut d'Astrophysique de Paris, UMR 7095, CNRS, 98 bis boulevard Arago, F-75014 Paris, France}
\affil{$^{10}$ Institut Lagrange de Paris, Sorbonne Universites, 98 bis Boulevard Arago, 75014 Paris, France}
\affil{$^{11}$ SISSA, Via Bonomea 265, 34136 Trieste, Italy}
\affil{$^{12}$ INFN - Istituto Nazionale di Fisica Nucleare, Sezione di Trieste, Via Bonomea 265, 34136 Trieste, Italy}
\affil{$^{13}$ INAF - Osservatorio Astronomico di Trieste, via Tiepolo 11, I-34143 Trieste, Italy}
\affil{$^{14}$ IFPU - Institute for Fundamental Physics of the Universe, Via Beirut 2, 34014 Trieste, Italy}
\affil{$^{15}$ Department of Physics, Carnegie Mellon University, Pittsburgh, PA 15213, USA}
\altaffiltext{*}{abayer@berkeley.edu}
\altaffiltext{$\dagger$}{fvillaescusa@flatironinstitute.org}

\begin{abstract}
We quantify the information content of the non-linear matter power spectrum, the halo mass function, and the void size function, using the \textit{Quijote} $N$-body simulations. We find that these three statistics exhibit very different degeneracies amongst the cosmological parameters, and thus the combination of all three probes enables the breaking of degeneracies, in turn yielding remarkably tight constraints. 
We perform a Fisher analysis using the full covariance matrix, including all auto- and cross-correlations, finding that this increases the information content for neutrino mass compared to a correlation-free analysis.
The multiplicative improvement of the constraints on the cosmological parameters obtained by
combining all three probes compared to using the power spectrum alone are: 137, 5, 8, 20, 10, and 43, for $\Omega_m$, $\Omega_b$, $h$, $n_s$, $\sigma_8$, and $M_\nu$, respectively.
The marginalized error on the sum of the neutrino masses is $\sigma(M_\nu) = 0.018 \,{\rm eV}$ for a cosmological volume of $1\,(h^{-1}{\rm Gpc})^3$, using $k_{\max}=0.5\,h{\rm Mpc}^{-1}$, and without CMB priors. We note that this error is an underestimate insomuch as we do not consider super-sample covariance, baryonic effects, and realistic survey noises and systematics.
On the other hand, it is an overestimate insomuch as our cuts and binning are suboptimal due to restrictions imposed by the simulation resolution.
Given upcoming galaxy surveys will observe volumes spanning $\sim 100\,(h^{-1}{\rm Gpc})^3$, this presents a promising new avenue to measure neutrino mass without being restricted by the need for accurate knowledge of the optical depth, which is required for CMB-based measurements.
Furthermore, the improved constraints on other cosmological parameters, notably $\Omega_m$, may also be competitive with CMB-based measurements.\\

\end{abstract} 


%%%%%%%%%%%%%%%%%%%%%%%%%%%%%%%%%%%%%%%%%%%%
\section{Introduction}
\label{sec:introduction}
\setcounter{footnote}{0} 

%The goal of upcoming cosmological missions like DESI\footnote{https://www.desi.lbl.gov}, Euclid\footnote{https://www.euclid-ec.org}, LSST\footnote{https://www.lsst.org}, PFS\footnote{https://pfs.ipmu.jp/index.html}, SKA\footnote{https://www.skatelescope.org}, and WFIRST\footnote{https://wfirst.gsfc.nasa.gov/index.html} is to provide answers to fundamental questions like what is the nature of dark energy? and what are the neutrino masses? To respond to these questions two ingredients are needed: 1) data from observations and 2) a theory prediction for the observables.

High-precision measurements of large-scale structure from upcoming cosmological surveys, such as
%DESI \citep{collaboration2016desi}, LSST  \citep{LSST_Science_Book}, Euclid \citep{EuclidDefn, Euclid, Euclid2}, eBOSS \citep{eBOSS_Dawson_2016}, WFIRST \citep{spergel2013widefield}, SKA \citep{SKA_Godfrey_2012}, and PFS \citep{Takada_2014}
DESI\footnote{\url{https://www.desi.lbl.gov}},
Euclid\footnote{\url{https://www.euclid-ec.org}}, PFS\footnote{\url{https://pfs.ipmu.jp/index.html}}, Roman Space Telescope\footnote{\url{https://wfirst.gsfc.nasa.gov/index.html}}, Vera Rubin Observatory\footnote{\url{https://www.lsst.org}}, SKA\footnote{\url{https://www.skatelescope.org}}, and SPHEREx\footnote{\url{https://www.jpl.nasa.gov/missions/spherex}}, 
are expected to revolutionize our understanding of fundamental physics, for example, by measuring neutrino mass. % and the nature of dark energy.
To fully realize the potential of these surveys, an urgent task is to determine the key observables that can maximize the scientific return. For Gaussian density fields, the answer is well known: the power spectrum, or equivalently, the correlation function, is the statistic that completely characterizes the field. Therefore, on large scales and at high redshift, where the density fluctuation in the Universe resembles a Gaussian field,  the power spectrum encapsulates all the information. 

However, at low redshift and on small scales, non-linear gravitational evolution moves information from the power spectrum into higher-order moments. It is currently ill-understood which observable(s) will allow  retrieval of the maximum information in the non-linear regime. For instance, it has been shown that for non-Gaussian fields, all clustering information may not be embedded in the infinite N-point statistics \citep{Carron_2011, Carron_2012}. Since the number of modes increases rapidly by going to small scales, it is expected that the amount of information will also increase by considering observables in the mildly to fully non-linear regime. While the amount of information, at least for some parameters, may saturate in the power spectrum \citep{Rimes_2005, quijote} \citep[see however][]{Blot_2016}, many authors have shown that other statistics contain complementary information  \citep[see, e.g.][]{Takada_2004, Sefusatti_2006, Berge_2010, Kayo_2013, Schaan_2014, Liu2015,Liux2015,Kacprzak2016,Shan2018,Martinet2018, Hahn_2020, hahn2020constraining, Dai_2020, Uhlemann_2020, Allys_2020, Gualdi_2020,Harnois-Deraps2020, Arka_2020, Massara_2020}.
%\jl{I only added higher-order stats from observations.}
% other higher-order stats paper, collected from 2020 snowmass white paper: 
%Jain2000b,Marian2009,Maturi2010,Yang2011,Marian+2013,Liu2015,Liux2015,Lin&Kilbinger2015a,Lin&Kilbinger2015b,Kacprzak2016,Peel2018,Shan2018,Martinet2018,Li2019,ajani2020,zuercher2020,Coulton2020,zuercher2020,Kratochvil2012,Shirasakiyoshida2014,Petri2013,Petri2015,Marques2019,Bernardeau1997,Hui1999,vanWaerbeke2001, Takada2002,Zaldarriaga2003,Kilbinger2005,Petri2015,Peel2018,Cheng2020

%In this paper we introduce the \textit{Big Covariance Project}\ab{?}, an enterprise aimed at quantifying the information embedded into different cosmological observables.

%In this paper we perform a Fisher analysis use the \textit{Quijote} simulations: a suite of 23000 $N$-body simulations for 16 different cosmologies expanding six cosmological parameters: $\Omega_m$, $\Omega_b$, $h$, $n_s$, $\sigma_8$ and $M_\nu$ \citep{quijote}.

In this paper we quantify the information embedded in the non-linear matter power spectrum, the halo mass function (HMF), and the void size function (VSF). 
We apply the Fisher formalism using a subset of the \textit{Quijote} simulations \citep{quijote}, comprising of 23,000 $N$-body simulations for 16 different cosmologies spanning six cosmological parameters: $\Omega_m$, $\Omega_b$, $h$, $n_s$, $\sigma_8$, and $M_\nu$.
We study the information that these probes contain individually and when combined together, showing how the combination of these three statistics breaks degeneracies amongst the cosmological parameters, in turn setting very tight constraints.
We consider the effects of both the auto-correlation for each probe and the cross-correlation between different probes when computing the total information content.
A simpler, theoretical, treatment combining cluster and void abundances has been studied by \cite{Sahl_n_2019}.

Of particular interest in this work are constraints on the sum of the neutrino masses $M_\nu \equiv \sum_\nu m_\nu$.
The first evidence for neutrino mass came from oscillation experiments \citep{SuperK, SNO, KamLAND, K2K, DayaBay},
%, however, such experiments are only sensitive to 
which measured the difference in the squares of the masses of the three neutrino mass eigenstates. The best-fit results obtained from a joint analysis of oscillation experiments are $\Delta m_{21}^2 \equiv m_2^2 - m_1^2 \simeq 7.55 \times 10^{-5} {\rm eV}^2$ from solar neutrinos, and $|\Delta m_{31}^2| \equiv |m_3^2 - m_1^2| \simeq 2.50 \times 10^{-3} {\rm eV}^2$ from atmospheric neutrinos \citep{3sig_nu}. Since atmospheric neutrino experiments only probe the magnitude of the mass difference, there are two possibilities for the neutrino mass hierarchy: $\Delta m_{31}^2 > 0$, known as the normal hierarchy, or $\Delta m_{31}^2 < 0$, known as the inverted hierarchy. This gives a lower bound on the sum of the neutrino masses of $M_\nu \gtrsim 0.06 {\rm eV}$ for the normal hierarchy, or $M_\nu \gtrsim 0.1 {\rm eV}$ for the inverted hierarchy. The current tightest upper bound on the effective electron neutrino mass from particle experiments is obtained  by the KATRIN $\beta$-decay experiment, $m_{\nu_{e}}^{\rm eff} \lesssim 1.1 {\rm eV}$ \citep{Aker_2019}~\footnote{Single $\beta$-decay experiments do not directly measure the neutrino mass sum, but rather the effective mass of electron neutrinos. In the quasi-degenerate regime where the eigenmasses $m_i>0.2$~eV ($i=1,2,3$),  the three eigenmasses are the same to better than 3\%, and hence $m_{\nu_{e}}^{\rm eff} \approx 1/3 M_\nu$.}.

Neutrinos also play an important role in the Universe's history, as the presence of massive neutrinos both shifts the time of matter-radiation equality and suppresses the growth of structure on small scales. 
%due to their high thermal velocities, distinguishing them from relatively slow cold dark matter (CDM) \citep{Doroshkevich_nu, Hu_1998, Eisenstein_1999, lesgourgues_mangano_miele_pastor_2013}; measuring this suppression enables determination of neutrino mass via cosmological structure formation.
%
Measuring these effects enables determination of neutrino mass via cosmology, providing a  complementary probe to particle physics \citep{Doroshkevich_nu, Hu_1998, Eisenstein_1999, lesgourgues_mangano_miele_pastor_2013}.
While the effects of neutrinos on linear (i.e.~relatively large) scales are well understood theoretically, understanding the effects on non-linear (i.e.~relatively small) scales is an active field of research. There are numerous approaches to obtain theoretical predictions of the non-linear effects of neutrinos, with varying computational efficiency 
\citep[see, e.g.][]{Saito_2008, Brandbyge_2009, Brandbyge_2010, shoji2010massive, Viel_2010, Ali_Ha_moud_2012, Bird_2012, hybrid, Costanzi_2013, Villaescusa_Navarro_2014, Villaescusa_Navarro_2018, Castorina_2014, Castorina_2015, Arka_2016, Archidiacono_2016, Carbone_2016, Upadhye_2016, Adamek_2017, Emberson_2017, Inman_2017, senatore2017effective, Yu_2017, Arka_2018, Liu2018MassiveNuS:Simulations, Dakin_2019, chen2020line, chen2020cosmic, Bayer_2021_fastpm}.
%In this work we use the \textit{Quijote} simulations \citep{quijote}, which have been specifically designed to produce the data necessary for a Fisher analysis for a generic statistic in the non-linear regime.

The current best constraints on $M_\nu$ arise by considering the cosmic microwave background (CMB) and combining it with other cosmological probes.
Assuming a $\Lambda$ cold dark matter ($\Lambda{\rm CDM}$) cosmological model, the upper bound on the neutrino mass from the Planck 2018 CMB temperature and polarization data is $M_\nu < 0.26 {\rm eV}$ (95\% CL) \citep{collaboration2018planck}. When combined with baryonic acoustic oscillations (BAOs) a more stringent bound of $M_\nu < 0.13 {\rm eV}$ (95\% CL) is obtained. Further combining with CMB lensing gives $M_\nu < 0.12 {\rm eV}$ (95\% CL). 
%Allowing more flexibility in the cosmological model, such as letting the spectral index 
%of inflationary perturbations
%run and considering a varying dark energy equation of state, can increase this upper bound to $M_\nu < 0.52 {\rm eV}$ (95\% CL) \citep{Valentino_2020}.
%
% These constraints are expected to improve by combining CMB data on large scales
% with clustering/lensing data on small scales and low redshifts, where the suppression of
% power by neutrinos is strongest \citep{Brinckmann_2019}. However, 

A major limiting factor of current cosmological constraints is that CMB experiments measure the combined
quantity $A_s e^{-2\tau}$, where $A_s$ is the amplitude of scalar perturbations and $\tau$ is the optical depth of reionization. Hence, accurate determination of $\tau$ is imperative to obtaining tight constraints when combining CMB with clustering/lensing \citep{Allison_2015, Liu_2016, Archidiacono_2017, Byeonghee_tau, Brinckmann_2019}. Most upcoming ground-based
CMB experiments, such as Simons Observatory and CMB-S4, will not observe scales larger than $\ell \sim 30$, and will therefore be unable to directly constrain $\tau$ \citep{CMBS4}. %Furthermore, CMB lensing typically shows strong degeneracies amongst cosmological parameters \ab{citation?}.
Planck currently provides the best constraint of $\tau = 0.054 \pm 0.007$, with large improvements expected from the ongoing CLASS experiment \citep{Watts_2018} and the upcoming LiteBIRD \citep{LiteBIRD} space mission.
Furthermore, future radio 21cm and, e.g., near-infrared/optical galaxy observations will provide new information on the optical depth which would also help improve the constraints form the CMB \citep{Liu_2016,Brinckmann_2019}.
 
Before significant progress will be made in measuring $\tau$, improved measurements of $M_\nu$ are expected from galaxy surveys such as DESI, LSST, and Euclid. These surveys will measure fluctuations on non-linear scales with unprecedented precision. 
There is thus much motivation to explore other probes of neutrino mass, beyond the traditional 2-point clustering. 
%, with upcoming galaxy surveys predicted to provide much more precise measurements of neutrino mass \citep{Font_Ribera_2014,SKA_Zhang_2020}. For example, DESI and LSST forecast constraints of order  $\sim 0.02{\rm eV}$, 
%thus the minimal neutrino mass should be detectable at the $\sim 3 \sigma$ level. Similar levels of accuracy are expected from CMB experiments when combined with BAO measurements from DESI. Having multiple independent probes of neutrino masses will allow not only a tighter constrain on the neutrino masses, but also a more robust one that will be less prone to systematics of each individual probe.
By adding probes such as the halo and void abundances, we demonstrate that it is possible to break the strong degeneracy between $M_\nu$ and $\sigma_8$ usually seen in 2-point clustering constraints \citep[see, e.g.][]{Villaescusa_Navarro_2018}. 
In turn, this gives tight constraints on neutrino mass, and in fact all cosmological parameters, potentially without the need for including CMB priors. In addition to improved constraints,  having multiple independent probes of neutrino masses will allow for more robust controls of systematics.

%The Hubble parameter, $h$, is also of particular interest, due to the well-known Hubble tension. Observations of supernovae, cepheids \citep{Riess_2019, Reid_2019}, and strong-lensing time delays \citep{Wong_2019} find a Hubble constant of around $H_0 \simeq 74 {\rm km/s/Mpc}$, while measurements of the CMB \citep{collaboration2018planck} and various probes of large-scale structure \citep{Abbott_2018, Cuceu_2019,Sch_neberg_2019,Ivanov_2020_BOSS,d_Amico_2020_EFT,Colas_2020,Philcox_2020} find $H_0 \simeq 67 {\rm km/s/Mpc}$. There is much debate as to whether this is due to systematics or a hint of new physics \citep[see, e.g.][for reviews]{Knox_2020,efstathiou2020lockdown}. %Typical error bars on $H_0$ are of order 1\% and there is much interest in finding new ways to tighten these constraints to help explain the Hubble tension.
%Constraints on $H_0$ from new probes can potentially shed light on the origin of this discrepancy. 

%JL: I removed this paragraph since it does not add much information here
%In this paper we focus our attention on the matter field, using the matter power spectrum and identifying voids in the underlying matter field in real-space. We will discuss in the conclusions how we expect our constraints and conclusions to change depending on these and other assumptions made in this paper, and how we plan to address them in subsequents works.

The paper is organized as follows. We first review the \textit{Quijote} simulations in Section~\ref{sec:Simulations}. The Fisher formalism used to quantify the information content on the different observables is described in Section~\ref{sec:Fisher}. We explain how the matter power spectrum, halo mass function, and void size function are obtained in Section~\ref{sec:probes}. We show the results of our analysis in Section~\ref{sec:Results}. Finally, we conclude in Section~\ref{sec:Conclusions}.

%%%%%%%%%%%%%%%%%%%%%%%%%%%%%%%%%%%%%%%%%%%%

\section{Simulations}
\label{sec:Simulations}

\begin{table*}
\begin{center}
\renewcommand{\arraystretch}{1.25}
\resizebox{0.85\textwidth}{!}{\begin{tabular}{ c | c  c  c  c  c  c | c | c }
\toprule
    \multicolumn{9}{c}{Quijote Simulations} \\ [3pt] \toprule
Name & $\Omega_m$ & $\Omega_b$ & $h$ & $n_s$ & $\sigma_8$ & $M_\nu$(eV) & ICs & Realizations \\[2pt]
\hline
\hline
    Fiducial 	    & 0.3175 & 0.049 & 0.6711 & 0.9624 & 0.834 & 0.0 & 2LPT & 15,000 \\ 
    Fiducial ZA     & 0.3175 & 0.049 & 0.6711 & 0.9624 & 0.834 & 0.0 & Zel'dovich& 500 \\ 
    $\Omega_m^+$    & \underline{0.3275} & 0.049 & 0.6711 & 0.9624 & 0.834 & 0.0 & 2LPT & 500 \\ 
    $\Omega_m^-$    & \underline{0.3075} & 0.049 & 0.6711 & 0.9624 & 0.834 & 0.0 & 2LPT & 500 \\ 
    $\Omega_b^{++}$ & 0.3175 & \underline{0.051} & 0.6711 & 0.9624 & 0.834 & 0.0 & 2LPT & 500 \\ 
    $\Omega_b^{--}$ & 0.3175 & \underline{0.047} & 0.6711 & 0.9624 & 0.834 & 0.0 & 2LPT & 500 \\ 
    $h^+$           & 0.3175 & 0.049 & \underline{0.6911} & 0.9624 & 0.834 & 0.0 & 2LPT & 500 \\ 
    $h^-$           & 0.3175 & 0.049 & \underline{0.6511} & 0.9624 & 0.834 & 0.0 & 2LPT & 500 \\ 
    $n_s^+$         & 0.3175 & 0.049 & 0.6711 & \underline{0.9824} & 0.834 & 0.0 & 2LPT & 500 \\ 
    $n_s^-$         & 0.3175 & 0.049 & 0.6711 & \underline{0.9424} & 0.834 & 0.0 & 2LPT & 500 \\ 
    $\sigma_8^+$    & 0.3175 & 0.049 & 0.6711 & 0.9624 & \underline{0.849} & 0.0 & 2LPT & 500 \\ 
    $\sigma_8^-$    & 0.3175 & 0.049 & 0.6711 & 0.9624 & \underline{0.819} & 0.0 & 2LPT & 500 \\
    $M_\nu^+$       & 0.3175 & 0.049 & 0.6711 & 0.9624 & 0.834 & \underline{0.1} & Zel'dovich & 500 \\ 
    $M_\nu^{++}$    & 0.3175 & 0.049 & 0.6711 & 0.9624 & 0.834 & \underline{0.2} & Zel'dovich & 500 \\ 
    $M_\nu^{+++}$   & 0.3175 & 0.049 & 0.6711 & 0.9624 & 0.834 & \underline{0.4} & Zel'dovich & 500 \\ %[3pt]
    \hline
\end{tabular}}
\caption{\label{table:sims}Characteristics of the subset of the \textit{Quijote} simulations used in this work. The fiducial cosmology contains 15,000 simulations, that are used to compute the covariance matrix. In the other cosmological models, one parameter is varied at a time, and these simulations are used to compute the numerical derivatives. The initial conditions of all simulations were generated at $z=127$ using 2LPT, except for the simulations with massive neutrinos and a copy of the fiducial cosmology, where the Zel'dovich approximation is used (see main text for further details). All realizations follow the evolution of $512^3$ CDM (+ $512^3$ Neutrino) particles in a box of size 1 $h^{-1}{\rm Gpc}$ down to $z=0$, with a gravitational softening length 50 $h^{-1}{\rm kpc}$. For massive neutrino simulations, we assume three degenerate neutrino masses. }
\end{center}
\end{table*}

We quantify the information content of different cosmological observables using the Fisher matrix formalism. We model the observables using the \textit{Quijote} simulations \citep{quijote}, a set of 23,000 $N$-body simulations that at a given redshift 
contain about 8 trillion ($8\times10^{12}$) particles over a total combined volume of 44,100 $(h^{-1}{\rm Gpc})^3$. Each simulation considers a box of size 1 $(h^{-1}{\rm Gpc})^3$. The simulation subset used in this work spans a total of 16 different cosmological models that have been designed to evaluate the two ingredients required to compute the Fisher matrix: (1) the covariance matrix of the observables and (2) the derivatives of the observables with respect to the cosmological parameters. Despite their larger computational cost than analytic approaches (e.g.~perturbation theory or the halo model), numerical simulations are more accurate into the fully non-linear regime and rely on fewer assumptions and approximations. 

We consider six cosmological parameters: $\Omega_m$, $\Omega_b$, $h$, $n_s$, $\sigma_8$, and $M_\nu$. The set of cosmological parameters is shown in Table \ref{table:sims}.
To evaluate the covariance matrix, we use the 15,000 simulations of the fiducial cosmology. We compute the derivatives by considering simulations where only one cosmological parameter is varied, with all others fixed. We use 1,000 simulations (500 pairs) for each derivative, with the exception of neutrino mass, where we use 1,500 (see below).

The initial conditions (ICs) were generated in all cases at $z=127$ using second-order Lagrangian perturbation theory (2LPT) for simulations with massless neutrinos, by rescaling the $z=0$ matter power spectrum using the scale-independent growth factor from linear theory. Because the 2LPT formalism has not yet been developed to account for massive neutrinos, the ICs for massive neutrino cosmologies adopt the Zel'dovich approximation with scale-dependent growth factors and rates, following \cite{Zennaro_2017}. 
For this reason there is also a `Fiducial (ZA)' class of simulations, which is identical to the fiducial simulations but with Zel'dovich ICs to match the $M_\nu$ simulations \citep[see][for further details]{quijote}; this enables accurate computation of derivatives with respect to $M_\nu$. Note that in the full \textit{Quijote} simulations there are two sets of $\Omega_b$ cosmologies; we use the $\Omega_b^{++}$ and $\Omega_b^{--}$ set too obtain smoother derivatives.

All simulations follow the evolution of $512^3$ dark matter particles down to $z=0$. The simulations with massive neutrinos also contain $512^3$ neutrino particles. The gravitational force tree for neutrinos is turned on at $z=9$. The gravitational softening for both dark matter and neutrinos is 50 $h^{-1}{\rm kpc}$ ($1/40$ of the mean interparticle distance). In this work, we consider redshift $z=0$ only. %We will discuss how we expect this choice to affect our findings in the conclusions.

%%%%%%%%%%%%%%%%%%%%%%%%%%%%%%%%%%%%%%%%%%%%
% \section{Methods}
% \label{sec:Methods}

% In this section we first outline the ingredients required to quantify the information content on cosmological observables using the Fisher analysis, and then describe the three different observables we use in this work.

\section{Fisher information}
\label{sec:Fisher} 

We use the Fisher matrix formalism \citep{Tegmark_96, Heavens_2007, Heavens_2010, Verde_2010} to calculate the information embedded in the non-linear matter power spectrum, the halo mass function and the void size function, individually and when combined. The Fisher matrix is defined as
\be
F_{ij} = -\left \langle \frac{\partial^2 \log \mathcal{L}}{\partial \theta_i \partial \theta_j}\right \rangle~,
\ee
where $\mathcal{L}$ is the likelihood and $\vec{\theta}$ is the vector representing the parameters of the model \citep{fisher_1925}. 
Under the assumption that the region around the maximum of the likelihood can be approximated as a multivariate normal distribution, one can write the Fisher matrix as
\begin{eqnarray}
F_{i j} &=& \frac{1}{2}\left[\frac{\partial \vec{O}}{\partial \theta_i}C^{-1}\frac{\partial \vec{O}^T}{\partial \theta_j} + \frac{\partial \vec{O}}{\partial \theta_j}C^{-1}\frac{\partial \vec{O}^T}{\partial \theta_i} \right]\nonumber\\
&&+ \frac{1}{2}{\rm Tr}\left[C^{-1}\frac{\partial C}{\partial \theta_i}C^{-1}\frac{\partial C}{\partial \theta_j} \right]~,
\label{Eq:Fisher}
\end{eqnarray}
where $\vec{O}$ is the vector with the values of the observables and $C$ is the covariance matrix. In order to avoid underestimating the errors, we follow \cite{Carron_2013} and neglect the dependence of the covariance on the cosmological parameters, by setting the last term of Eq.~\ref{Eq:Fisher} to zero. This is necessary when assuming a Gaussian likelihood. Note that we use Greek (Latin) characters to index observables (parameters). 

In this work, the observables and parameters are given by
\begin{align*}
\vec{O}&=\{P_m(k_1),...,P_m(k_A), \mathcal{H}(M_1),...,\mathcal{H}(M_B),\\
&\textcolor{white}{.......................................}        \mathcal{V}(R_1),...,\mathcal{V}(R_D) \},\\
\vec{\theta} &= \{ \Omega_m, \Omega_b, h, n_s, \sigma_8, M_\nu\}
\end{align*}
respectively, 
where $P_m(k)$ is the matter power spectrum at wavenumber $k$, $\mathcal{H}(M)$ is the halo mass function at mass $M$, and $\mathcal{V}(R)$ is the void size function at radius $R$. Note there are a total of $A$, $B$, and $D$ bins for the matter power spectrum, the halo mass function, and the void size function respectively, giving a total dimensionality of $A+B+D$.

%We quantify the information content on the different observables by making use of the Cramer-Rao bound, that states the the variance of the optimal unbiased estimator depends on the Fisher matrix as
%\begin{equation}
%    \delta \theta_i \geq \sqrt{(F^{-1})_{i i}}~.
%\end{equation}

We quantify the information content by considering the marginalized error on the cosmological parameters,
\begin{equation}
    \sigma (\theta_i) \equiv \sqrt{(F^{-1})_{i i}}~,
    \label{eq:marg_err}
\end{equation}
which is a lower bound.

\subsection{Covariance matrix}
\label{subsec:Covariance}

We estimate the covariance matrix using the $N_{\rm cov}=$ 15,000 simulations of the fiducial cosmology as 
\be
C_{\alpha\beta}=\left\langle\left( O_\alpha - \langle O_\alpha \rangle \right)\left( O_\beta - \langle O_\beta \rangle \right)\right\rangle,
\label{eq:covm}
\ee
where $\langle\rangle$ denotes the mean over simulations. 
This is the largest number of simulations used for covariance estimation to date. We have verified that our combined results are converged even with half of the simulations. We show the results of our convergence tests in Appendix \ref{app:robustness}.

\subsection{Derivatives}
\label{subsec:Derivatives}

For the cosmological parameters $\Omega_m$, $\Omega_b$, $h$, $n_s$, and $\sigma_8$,
we approximate the derivatives using a central difference scheme centered on the fiducial cosmology,
\be
\frac{\partial \vec{O}}{\partial \theta_i}\simeq\frac{\vec{O}(\theta_i+\delta \theta_i)-\vec{O}(\theta_i-\delta \theta_i)}{2\theta_i}.
\label{eqn:CDS}
\ee
Note that only the value of the $i^{\rm th}$ cosmological parameter is perturbed about its fiducial value, $\theta_i$, while the values of all other parameters are held fixed.
The error of this approximation is $\mathcal{O}(\delta \theta_i^2)$. 

For neutrinos we cannot use Eq.~\ref{eqn:CDS} because the fiducial model has massless neutrinos, so $\vec{O}(\theta_i-\delta \theta_i)$ would correspond to a cosmology with negative neutrino mass. We thus compute the derivatives for neutrinos using a 
second-order
forward difference scheme,
%either with the first-order forward derivative,
%\be
%\frac{\partial \vec{O}}{\partial M_\nu}\simeq\frac{\vec{O}(M_\nu+\delta M_\nu)-\vec{O}(M_\nu)}{\delta M_\nu},
%\label{eq:error_nu1}
%\ee
%with an error of order $\mathcal{O}(\delta M_\nu)$  on the derivative estimation  or by combining the first-order and higher-order derivatives, 
\be
\frac{\partial \vec{O}}{\partial M_\nu}\simeq\frac{-3\vec{O}(M_\nu+2\delta M_\nu)+4\vec{O}(M_\nu+\delta M_\nu)-3\vec{O}(M_\nu)}{2\delta M_\nu},
\label{eq:error_nu2}
\ee
which has error $\mathcal{O}(\delta M_\nu^2)$. %To achieve higher precision, we adopt Eq.~\ref{eq:error_nu2}. However, we use Eq.~\ref{eq:error_nu1} for the void size function as this particular probe is shown to be more prone to numerical noises when using Eq.~\ref{eq:error_nu2}.
We exclusively use the $M_\nu^{++}$ and $M_\nu^{+++}$ cosmologies in Eq.~\ref{eq:error_nu2} throughout this work.

We use a total of $N_{\rm der}=$ 1,000 (500+500) simulations to compute derivatives when using Eq.~\ref{eqn:CDS}, and 1,500 when using Eq.~\ref{eq:error_nu2}. In Appendix \ref{app:robustness} we show that our results are robust and converged with this number of simulations. We also give evidence of robustness with respect to the choice of finite difference scheme for $M_\nu$.

\section{Cosmological probes}
\label{sec:probes}

In this section we outline the cosmological observables considered in this work: the matter power spectrum, the halo mass function, and the void size function. 

\subsection{Matter power spectrum}
\label{subsec:Pk}

The first observable we study is the matter power spectrum. For each realization, the density field is computed by depositing particle masses to a regular grid using the cloud-in-cell mass assignment scheme. In simulations with massive neutrinos we consider both CDM and neutrino particles when constructing the density field. The density contrast field, $\delta(\vec{x})=\rho(\vec{x})/\bar{\rho}-1$, is then Fourier transformed and the power spectrum is computing by averaging $|\delta(\vec{k})|^2$ over spherical bins in $|k|$. The size of each bin is equal to the fundamental frequency, $2\pi/L$, where $L=1\,h^{-1}{\rm Gpc}$ is the simulation box size. 

\begin{figure}
\begin{center}
\includegraphics[width=0.475\textwidth]{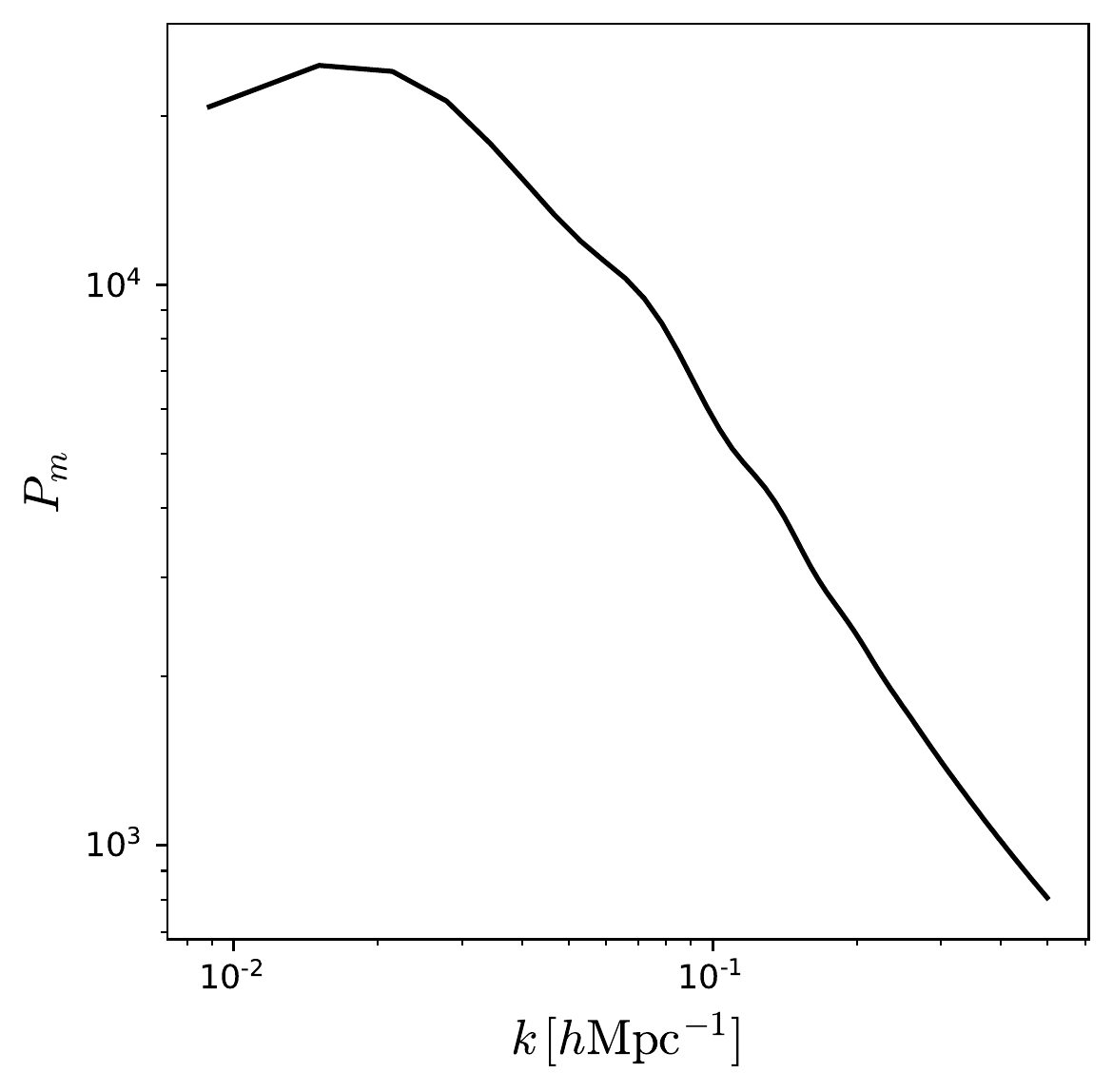}
\caption{\label{fig:Pm}The matter power spectrum for the fiducial cosmology.
}
\end{center}
\end{figure}

A grid with $1024^3$ cells is used, which is large enough to avoid aliasing effects on the scales of interest for this work. In our analysis we consider wavenumbers up to $k_{\rm max} = 0.5~h{\rm Mpc}^{-1}$, using 79 bins. This choice of $k_{\rm max}$ is based on the fact that the clustering of the simulations is converged at this scale for this mass resolution \citep[see][]{quijote}. We will however show that using a larger $k_{\rm max}$ would likely lead to even tighter constraints than the ones we report. We show the power spectrum for the fiducial cosmology in Fig.~\ref{fig:Pm}.

\subsection{Halo mass function}
\label{subsec:HMF}

The second observable we consider is the halo mass function (HMF). Dark matter halos are identified using the Friends-of-Friends algorithm \citep{FoF}, with a linking length $b=0.2$. The halo finder considers only the dark matter distribution, as the contribution of neutrinos to the total mass of a halo is expected to be negligible \citep{Paco_11, Paco_12, Ichiki_Takada_2011, LoVerde_2014}.

The halo mass function is defined as the comoving number density of halos per unit of (log) halo mass, $dn/d\ln M$. The mass of a halo is estimated as
\begin{equation}
    M = N m_p,
    \label{eq:MNm}
\end{equation}
 where $N$ is the number of dark matter particles in the halo and $m_p$ is the mass of a single dark matter particle. Note that in the \textit{Quijote} simulations, there are only dark matter and neutrino particles, i.e. dark matter particles represent the CDM+baryon fluid.  The mass of a dark matter particle is thus normalized according to $\Omega_{cb}$, such that
\begin{align}
    m_p 
    = \frac{V \rho_c}{N_p} \Omega_{cb}
    %&= \frac{V \rho_c}{N_p} \left( \Omega_m - \Omega_\nu \right) %\\
    &= \frac{V \rho_c}{N_p} \left( \Omega_m - \frac{M_\nu}{93.14 h^2} \right),
    \label{eq:mp}
\end{align}
where $V=L^3$ is the simulation volume, $N_p$ is the total number of dark matter particles in the simulation, and $\rho_c$ is the Universe's critical energy density at $z=0$. Thus $m_p = m_p(\Omega_m, M_\nu)$ is a cosmology dependent quantity, which induces noise when computing the derivatives of the HMF with respect to $\Omega_m$ or $M_\nu$ in a fixed \textit{mass} bin. This is because it is the \textit{number} of dark matter particles that is the fundamental constituent of the halo mass: a halo with a given number of particles will lie in the same \textit{number} bin for all cosmologies, whereas it may lie in a different \textit{mass} bin depending on the value of $m_p$. This noise can thus be avoided by instead working with bins of fixed particle \textit{number} by considering the derivative of the comoving number density of halos per unit (log) number of particles, $dn/d\ln N$. One can then transform these derivatives in bins of fixed $N$ to derivatives in bins of fixed $M$ to obtain the derivatives of the halo mass function. 

Using the shorthand $\mathcal{H}$ to denote the halo mass function, we now derive this transformation.
In practice, one measures the halo mass function for a fixed cosmology, thus working in logarithmic bins gives
\begin{equation}
    \mathcal{H} := \frac{dn}{d\ln M} = \frac{dn}{d\ln N},
\end{equation}
where it is understood that the derivative is taken with fixed cosmological parameters, $\vec{\theta}$.
Explicitly, one can think of the halo mass function as a function of the cosmological parameters and halo mass, $\mathcal{H}(\vec{\theta},M)$, or the cosmological parameters and number of particles, $\mathcal{H}(\vec{\theta},N)$. Thus the derivative of the HMF with respect to one of the cosmological parameters, $\theta$, while holding all other cosmological parameters, $\slashed{\theta}$, fixed can be written as 
%\begin{equation}
%    \frac{d\mathcal{H}}{d\theta} = \left(  \frac{\partial \mathcal{H}}{\partial\theta} \right)_{M} + \left( \frac{\partial \mathcal{H}}{\partial\ln M} \right)_{\vec{\theta}} \frac{d \ln M}{d \theta}
%\end{equation}
\begin{equation}
    \left( \frac{\partial \mathcal{H}}{\partial\theta} \right)_{\slashed{\theta}} = \left(  \frac{\partial \mathcal{H}}{\partial\theta} \right)_{M, \slashed{\theta}} + \left( \frac{\partial \mathcal{H}}{\partial\ln M} \right)_{\vec{\theta}} \left( \frac{ \partial \ln M}{\partial \theta} \right)_{\slashed{\theta}},
\end{equation}
or
%\begin{equation}
%    \frac{d\mathcal{H}}{d\theta} = \left(  \frac{\partial \mathcal{H}}{\partial\theta} \right)_{N} + \left( \frac{\partial \mathcal{H}}{\partial\ln N} \right)_{\vec{\theta}} \frac{d \ln N}{d \theta}.
%\end{equation}
\begin{equation}
    \left( \frac{\partial \mathcal{H}}{\partial\theta} \right)_{\slashed{\theta}} = \left(  \frac{\partial \mathcal{H}}{\partial\theta} \right)_{N,\slashed{\theta}} + \left( \frac{\partial \mathcal{H}}{\partial\ln N} \right)_{\vec{\theta}} \left( \frac{\partial \ln N}{\partial \theta} \right)_{\slashed{\theta}}.
\end{equation}
Equating these two equations and rearranging gives
%\begin{align}
%     \left( \frac{\partial \mathcal{H}}{\partial\theta} \right)_{M} 
%    &= 
%     \left( \frac{\partial \mathcal{H}}{\partial\theta}  \right)_{N}
%    +  \left( \frac{\partial \mathcal{H}}{\partial\ln N} \right)_{\vec{\theta}} 
%    \left[ \frac{d \ln N}{d \theta} - \frac{d \ln M}{d \theta} \right] \\
%    &=
%     \left( \frac{\partial \mathcal{H}}{\partial\theta}  \right)_{N}
%    - \left( \frac{\partial \mathcal{H}}{\partial\ln N} \right)_{\vec{\theta}} 
%    \frac{d \ln m_p}{d \theta}.
%\end{align}
\begin{align}
     \left( \frac{\partial \mathcal{H}}{\partial\theta} \right)_{M,\slashed{\theta}} 
    &= 
     \left( \frac{\partial \mathcal{H}}{\partial\theta}  \right)_{N,\slashed{\theta}}
    +  \left( \frac{\partial \mathcal{H}}{\partial\ln N} \right)_{\vec{\theta}} 
    \left[ \frac{\partial \ln N}{\partial \theta} - \frac{\partial \ln M}{\partial \theta} \right]_{\slashed{\theta}}\nonumber~\\
    &=
     \left( \frac{\partial \mathcal{H}}{\partial\theta}  \right)_{N,\slashed{\theta}}
    - \left( \frac{\partial \mathcal{H}}{\partial\ln N} \right)_{\vec{\theta}} 
    \left( \frac{\partial \ln m_p}{\partial \theta} \right)_{\slashed{\theta}},
    \label{eq:HMF_correction}
\end{align}
where Eq.~\ref{eq:MNm} was used in the final step.

The cosmology dependence of $m_p$ takes effect in the final term of Eq.~\ref{eq:HMF_correction}. There is only a difference between the fixed $N$ and fixed $M$ derivative of the HMF when $m_p$ depends on $\theta$, i.e., when $\theta \in \{\Omega_m, M_\nu\}$. Using Eq.~\ref{eq:mp}, one finds that
\begin{align}
    \frac{\partial \ln m_p}{\partial \Omega_m} 
    &=\frac{1}{\Omega_m}, 
    \label{eq:mp_Om}\\
    \frac{\partial \ln m_p}{\partial M_\nu} 
    &=\frac{1}{\Omega_{cb} 93.14 h^2},
    \label{eq:mp_Mnu}
\end{align}
where it is understood that all cosmological parameters apart from the one in the derivative are held fixed at their fiducial values.

Thus our procedure to compute derivatives of the HMF using Eq.~\ref{eq:HMF_correction} is as follows. We first bin the number of halos according to the number of dark matter particles they contain. We then compute the derivatives for each fixed-$N$ bin using the equations from Section \ref{subsec:Derivatives}, yielding the first term on the right-hand side of Eq.~\ref{eq:HMF_correction}. This will be sufficient for all cosmological parameters except for $\Omega_m$ and $M_\nu$, as these require a correction term to transform to fixed-$M$ bins due to the variation of $m_p$. The $\partial \mathcal{H}/\partial \ln N$ term can be computed 
via spline interpolation or
by using finite difference methods between the bins of the halo mass function of the fiducial cosmology. We have confirmed the stability of both approaches.
%, and find best results are achieved by fitting a smoothed spline through all points for all except the left-most bin, for which we use a forward difference scheme to reduce boundary effects. 
Finally, the derivative of $\ln m_p$ with respect to $\theta$ is computed using Eqs.~\ref{eq:mp_Om} and \ref{eq:mp_Mnu} evaluated at the fiducial values.

\begin{figure}
\begin{center}
\includegraphics[width=0.475\textwidth]{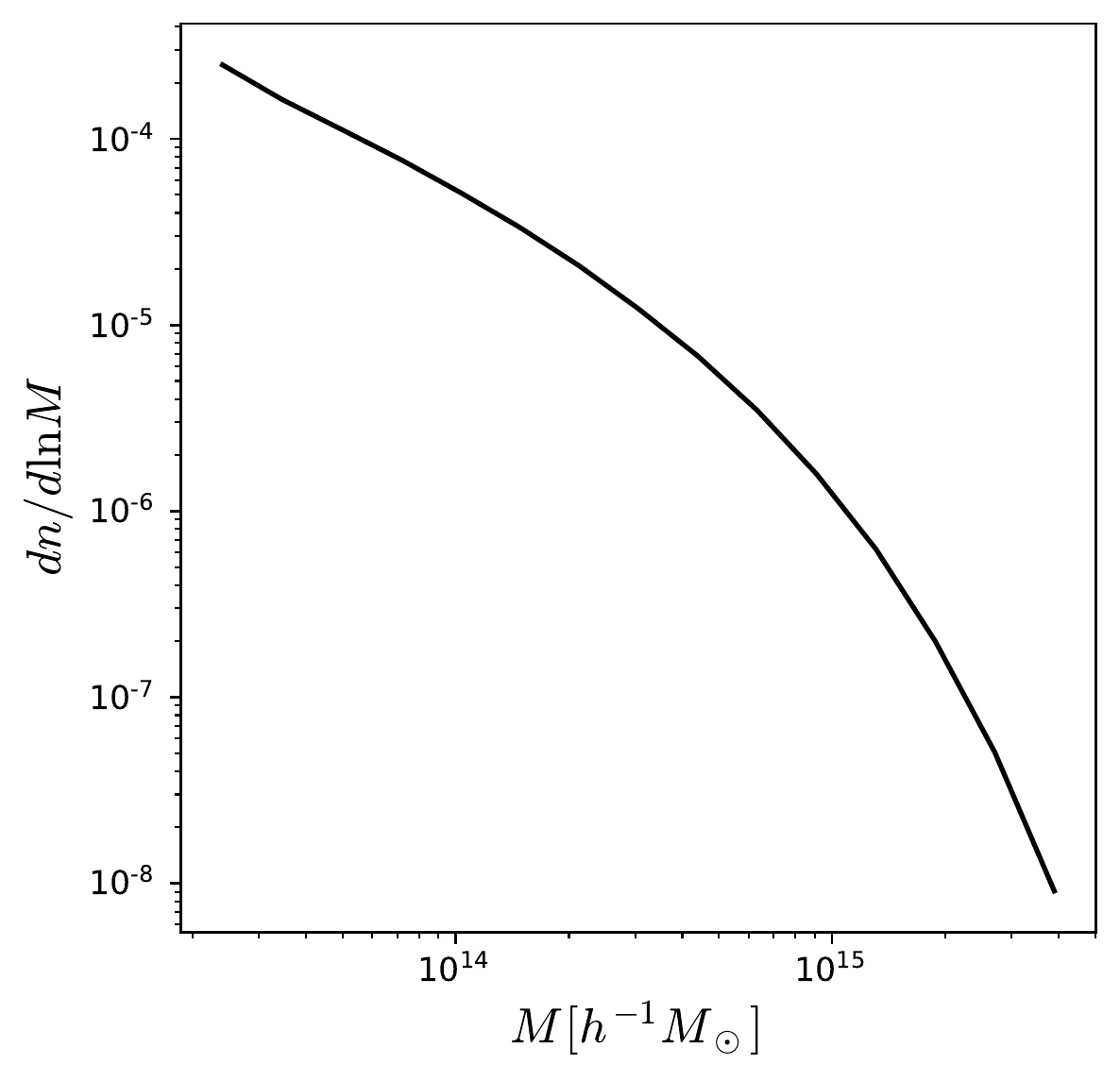}
\caption{\label{fig:HMF}The halo mass function for the fiducial cosmology.
}
\end{center}
\end{figure}

We consider halos with a number of dark matter particles between 30 and 7,000, using 15 logarithmically spaced bins. The corresponding halo mass range is approximately $2.0\times10^{13}$ 
%$h^{-1}M_\odot$ 
to $4.6\times10^{15}$ $h^{-1}M_\odot$.
As with the matter power spectrum, this choice of binning and cuts is made to ensure convergence of the derivatives based on the resolution and number of the simulations available. Hence, using more bins and/or a larger mass range would likely lead to stronger constraints than we report. We show the HMF for the fiducial cosmology in Fig.~\ref{fig:HMF}.

\subsection{Void size function}
\label{subsec:VSF}

We identify voids in the underlying matter field using a spherical void finding algorithm developed by \cite{Arka_2016}, which we now outline. We use a grid of resolution $768^3$ to look for voids --- this is slightly finer than the CDM grid resolution of $512^3$ to enable detection of small voids.
The density contrast field is then smoothed with a top-hat filter over a large-scale, $R=53.4\, h^{-1}$Mpc, which is a multiple of the grid spacing and is chosen to be bigger than the size of the largest void. Next, minima that are smaller than the threshold $\delta_{\rm th}=-0.7$ in the smoothed field are considered as voids with radius $R$, unless they overlap with existing voids. This procedure is then performed iteratively while decrementing $R$ by the grid spacing. 
In this work we use a threshold of $\delta_{\rm th}=-0.7$, but have checked that results are similar for $\delta_{\rm th}=-0.5$.

\begin{figure}
\begin{center}
\includegraphics[width=0.475\textwidth]{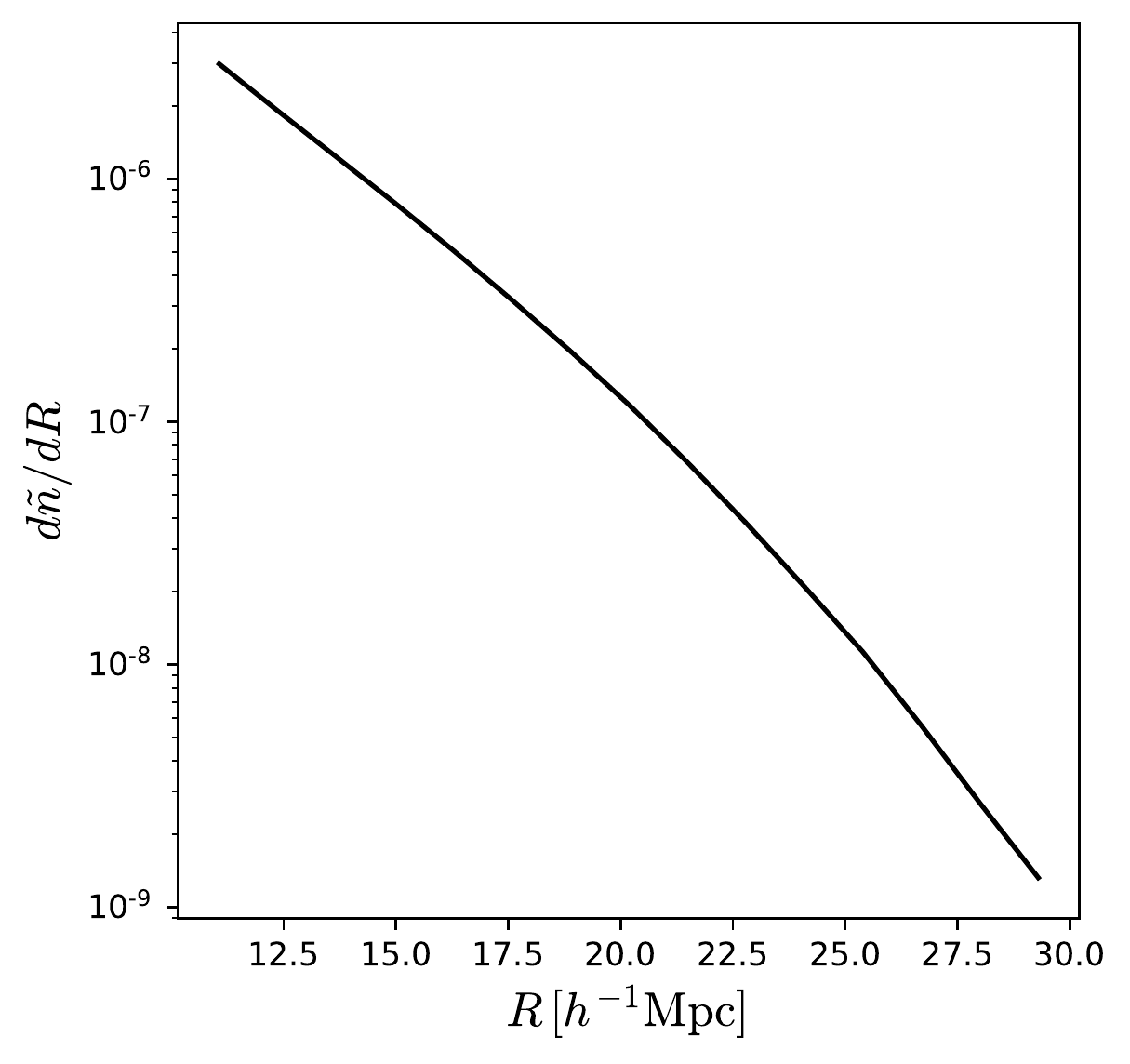}
\caption{\label{fig:VSF}The void size function for the fiducial cosmology.
}
\end{center}
\end{figure}

The void size function (VSF) is then computed as the comoving number density of voids per unit of radius, denoted $d\tilde{n}/dR$. Unlike the halo mass function, the VSF is not prone to the changes in particle mass, since the void finder operates directly in the same unit as the VSF.  The range of void sizes is limited by our resolution and the size of our simulated volume.
Having found the voids, we apply radius cuts of $R_{\rm min}=10.4$ and $R_{\rm max}=29.9$ $h^{-1}$Mpc, corresponding to 15 bins linear in $R$. 
As with the matter power spectrum and the halo mass function, this choice of binning and cuts is made to ensure convergence of the derivatives based on the resolution and number of the simulations available. Hence, using more bins and/or a larger range of void sizes may lead to stronger constraints than we report. We show the VSF for the fiducial cosmology in Fig.~\ref{fig:VSF}.

Investigation of the void size function, and void abundances, is a rich field that has shown promising theoretical work to match mocks \cite[see, e.g.][]{Platen_2008, Bos_2012, Sutter_2012, Jennings_2013, Pisani_2015, Paillas2017, Sahl_n_2019, Contarini_2019, Verza_2019}.

%%%%%%%%%%%%%%%%%%%%%%%%%%%%%%%%%%%%%%%%%%%%
\section{Results}
\label{sec:Results}

In this section we present the main results of this work.

\subsection{Full covariance of the probes}
\label{sec:fullcov}
\begin{figure}
\begin{center}
\includegraphics[width=0.47\textwidth]{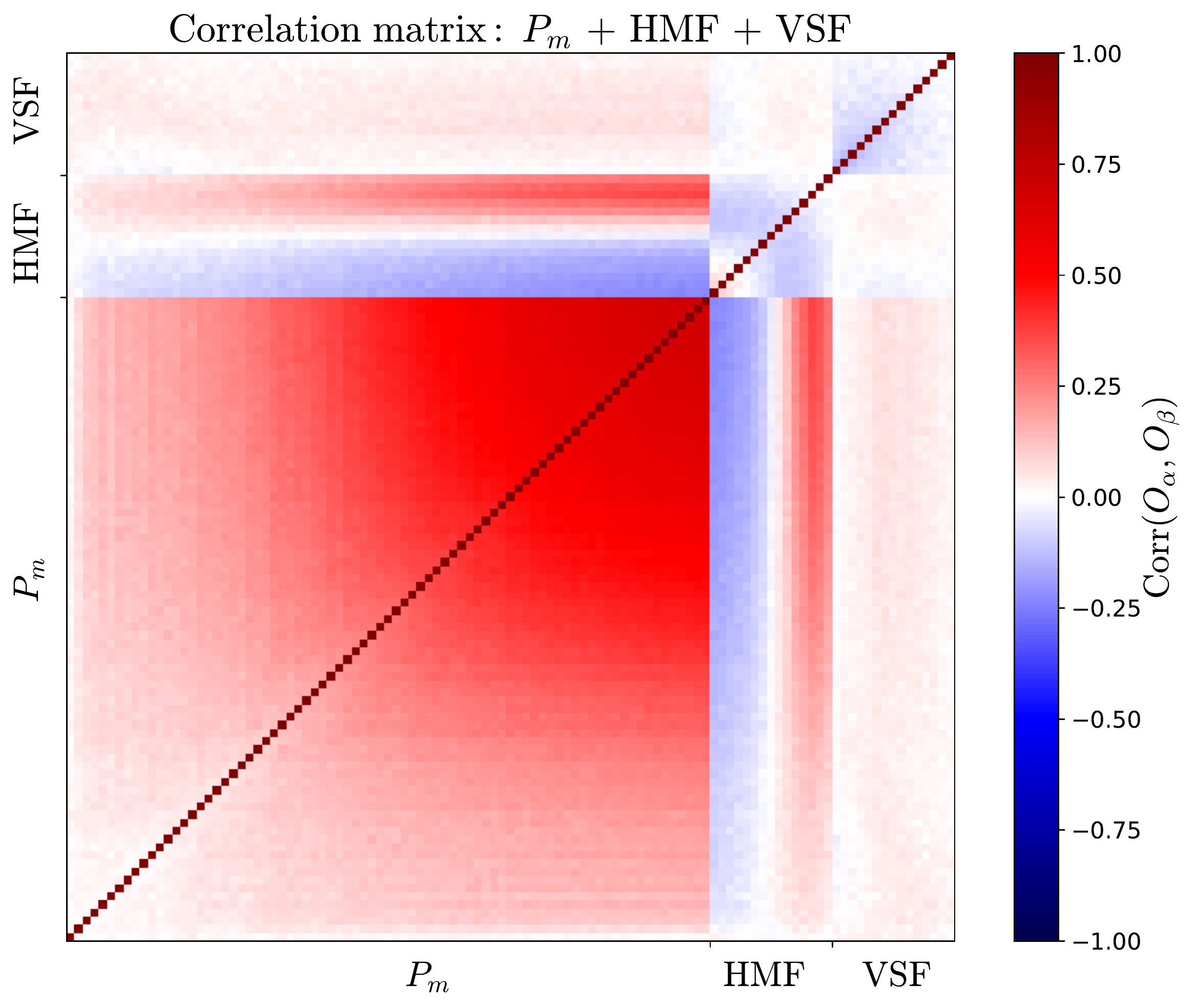}
\caption{Correlation matrix for the matter power spectrum ($P_m$, with 72 linear bins and $k_{\rm max}=0.5~h{\rm Mpc}^{-1}$), the halo mass function (HMF, 15 log bins between $2.0\times10^{13}$ and $4.6\times10^{16}~h^{-1}M_\odot$), and the void size function (VSF, 15 linear bins between 10.4 and 29.9 $h^{-1}{\rm Mpc}$), from bottom left to top right. Bin values increase from left to right for each probe. While the HMF shows clear off-block correlation with $P_m$, the VSF is somewhat independent from both $P_m$ and the HMF. 
%The correlation of the HMF and VSF are almost diagonal, with a small anti-correlation showing up in the off diagonal elements, except for small masses where the HMF shows correlation. The cross-correlation between the HMF and VSF is very small, suggesting the two probes to be essentially independent. Larger correlations show up between the HMF and $P_m$.
}
\label{fig:Covariance}
\end{center}
\end{figure}

\begin{figure*}
\begin{center}
\includegraphics[width=\textwidth]{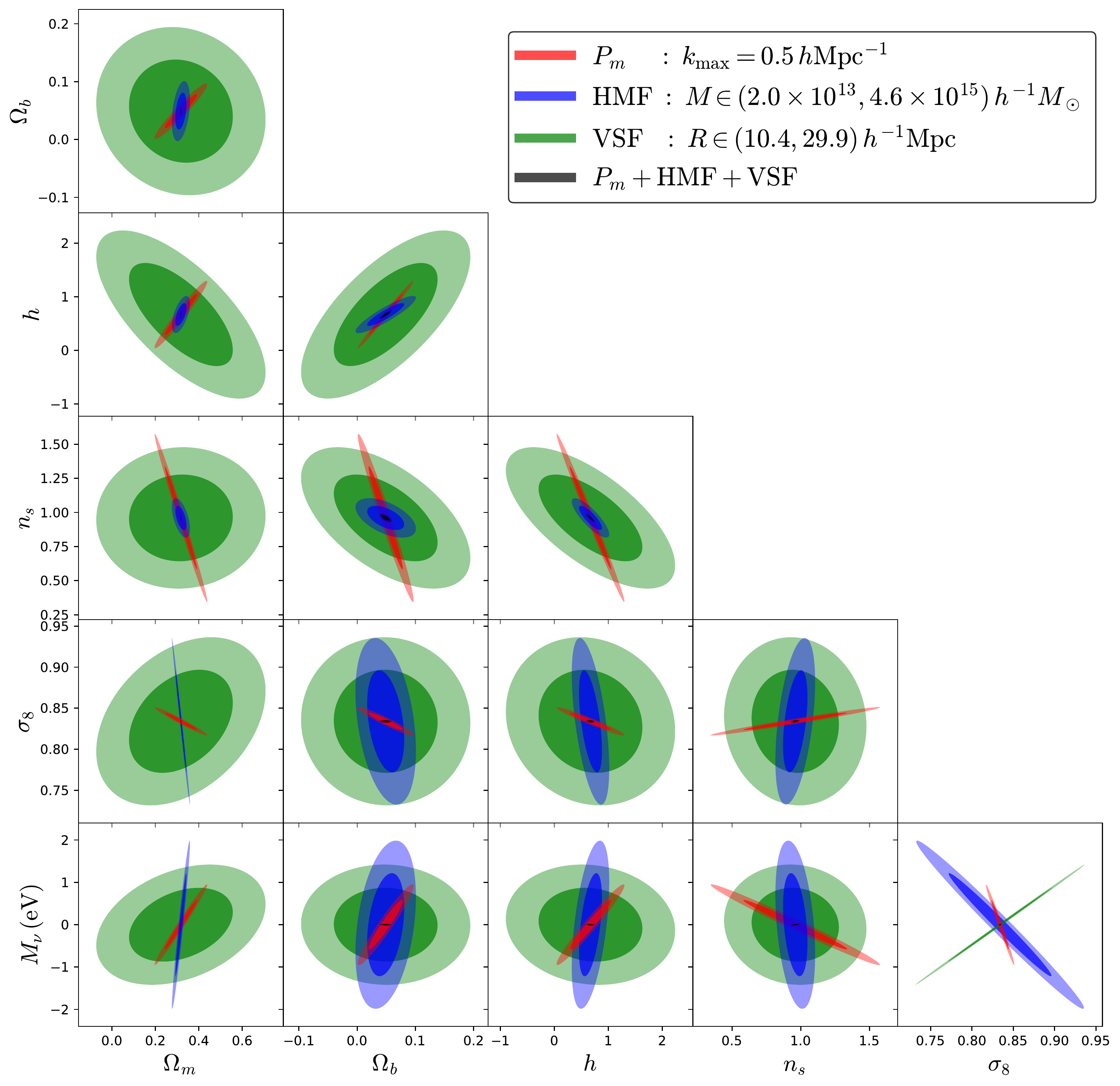}
\caption{\label{fig:Fisher_individual}68\% (darker shades) and 95\% (lighter shades) confidence contours for the cosmological parameters for the non-linear matter power spectrum ($P_m$, red), the halo mass function (HMF, blue), and the void size function (VSF, green). 
Due to the often different degeneracies of each probe, we obtain significantly tighter constraints when combining the three probes (black). We note that some contours extend into unphysical regions ($\Omega_b<0, h<0, M_\nu<0$): this is just a result of the Gaussian approximation associated with a Fisher analysis.}
\end{center}
\end{figure*}

\begin{figure}
\begin{center}
\includegraphics[width=0.475\textwidth]{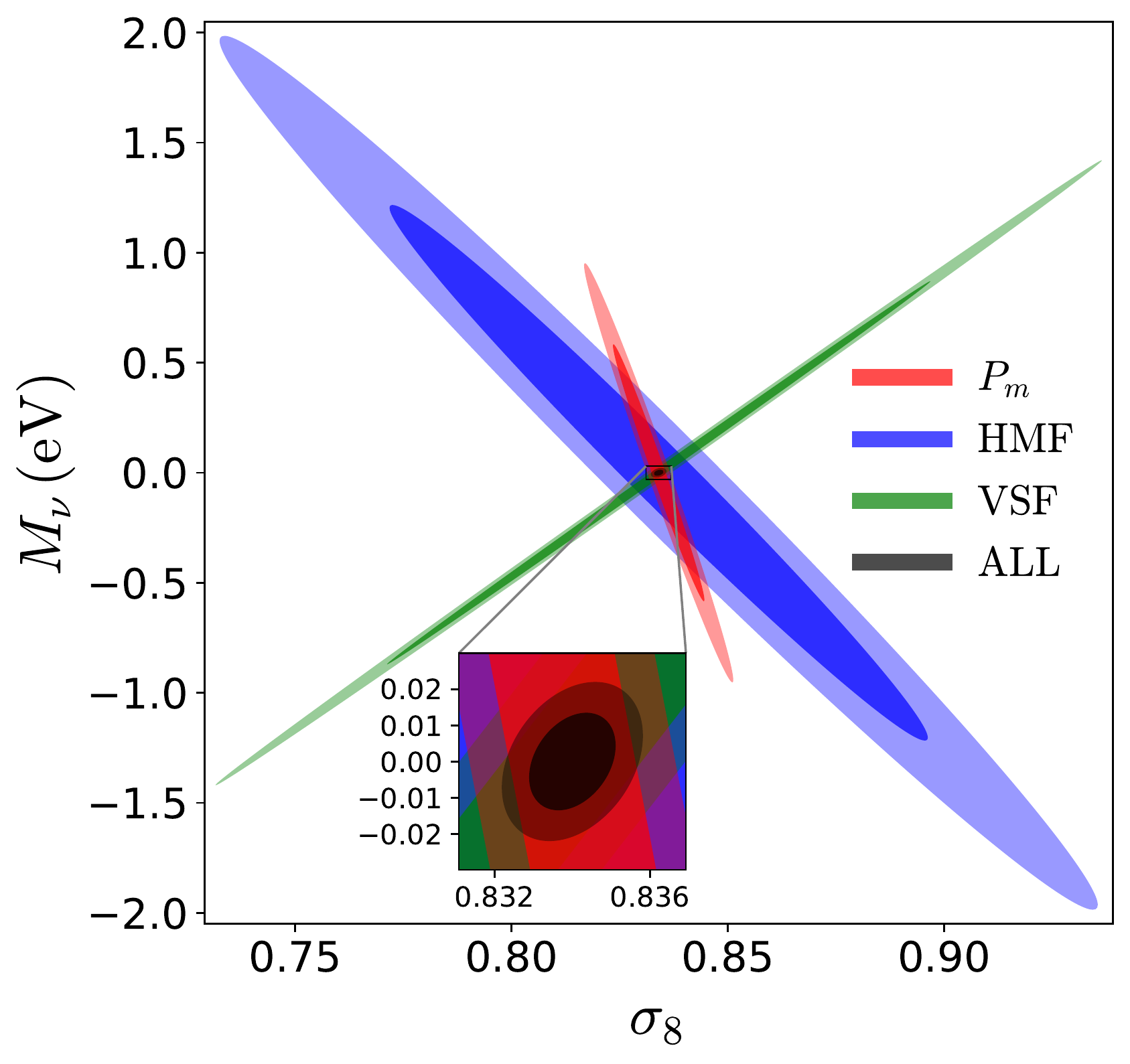}
\caption{\label{fig:Mnu-s8}The $M_\nu$--$\sigma_8$ plane from Fig.~\ref{fig:Fisher_individual}. We inset a zoom-in of the contour obtained by combining all three probes. The marginalized error on $M_\nu$ from $P_m$ alone is $0.77 {\rm eV}$, while the error after combining all three probes is $0.018 {\rm eV}$, corresponding to a factor $\sim 43$ improvement.
}
\end{center}
\end{figure}

In Fig.~\ref{fig:Covariance} we show the correlation matrix, defined as ${\rm Corr}(O_\alpha,O_\beta):=C_{\alpha\beta}/\sqrt{C_{\alpha\alpha}C_{\beta\beta}}$, where $C_{\alpha\beta}$ is the covariance matrix (Eq.~\ref{eq:covm}).
First we discuss the correlations for each individual probe (auto-correlations).
For the matter power spectrum (bottom-left region of Fig.~\ref{fig:Covariance}), we observe some well-known structures: the covariance is almost diagonal on large scales, while mode-coupling induces significant off-diagonal correlations on small scales. For the halo mass function (central region of Fig.~\ref{fig:Covariance}), the covariance matrix is almost diagonal,
%for all the considered halo masses, [$2.0\times10^{13}$, $4.6\times10^{15}$] $h^{-1}M_\odot$, 
with some small correlations between the different mass bins; the correlations are negative for heavy halos, but are positive for the lightest halos considered in this work. The covariance of the void size function (top-right region of Fig.~\ref{fig:Covariance}) is also almost diagonal, with the abundance of different void sizes slightly anti-correlated with nearby bins due to conservation of volume.

Next, we consider the correlations between different probes (cross-correlations).
The halo mass function shows an interesting correlation pattern with the matter power spectrum: the abundance of the more (less) massive halos shows a $\sim 20\%$ correlation (anti-correlation) with small scales of the matter power spectrum. Similar trends are seen between halos and large scales of the matter power spectrum, albeit at a weaker level.
On the other hand, voids can be seen to be somewhat independent of both the matter power spectrum and halos, as their cross-correlation is $\lesssim 5\%$ for all scales and masses.

As discussed in Section \ref{sec:Fisher}, we combine the covariance matrix with the numerically computed derivatives to calculate the Fisher matrix. 
The numerical derivatives and related numerical convergence tests are shown in Appendix \ref{app:robustness}. 

\subsection{Cosmological constraints}
\label{sec:cosmo_constraints}

We show the two-dimensional (2D) 68\% and 95\% confidence intervals obtained from our Fisher analysis for each individual probe, and the combination of all probes, in Fig.~\ref{fig:Fisher_individual}. 
The constraints on the parameters are not generally tight when considering any of three probes alone, because we adopt a conservative survey volume of 1 $(h^{-1}{\rm Gpc})^3$, which is significantly smaller than what is achievable by DESI, $\sim 10^2\,(h^{-1}{\rm Gpc})^3$.
%up to 6.4 $(h^{-1}{\rm Gpc})^3$~\citep{collaboration2016desi}. 

The three probes show different degeneracies and are sensitive to each parameter at different levels. For example, the halo mass function provides a relatively tight constraint on $\Omega_m$ when compared to the other two probes, as the halo mass function depends non-linearly on and is highly sensitive to $\Omega_m$~\citep[see, e.g.][]{Haiman2001}. The void size function provides weaker constraints than the other two probes on almost all parameters, except for $n_s$ compared to $P_m$. Naively, this is not surprising, considering the relatively smaller range of scales being probed by the void size function compared to the matter power spectrum. More information could probably be retrieved by using other void-related observables, such as the void-matter correlation function.

Because the degeneracies between parameters are often very different for each probe, %For instance, while $\Omega_m$ and $h$ are positively correlated for the matter power spectrum, they are anti-correlated for the void size function, and are almost uncorrelated for the halo mass function. It is thus naively 
it is expected that combining the probes will break the degeneracies and in turn yield significantly tighter constraints on the cosmological parameters than the individual probes do.
Indeed, the black ellipses in Fig.~\ref{fig:Fisher_individual} show the tight constraints obtained by combining the three probes. 
We emphasize that these constraints account for all the correlations between the different observables, i.e.~by using the full covariance matrix of Fig.~\ref{fig:Covariance}. 

The benefit of combining the three probes is particularly well demonstrated in the $M_\nu$--$\sigma_8$ plane. Because the combined constraints are too small to be visible in Fig.~\ref{fig:Fisher_individual}, we zoom in on this plane in Fig.~\ref{fig:Mnu-s8}.  
We find that, despite not being as powerful tools as $P_m$ in constraining $M_\nu$, the HMF and VSF both show degeneracies in different directions from that of $P_m$, which guarantees that constraints on the neutrino masses will be largely reduced by combining the three probes. 
In turn this helps break the well-known $M_\nu$--$\sigma_8$ degeneracy for the matter power spectrum. 
We note that the area of these confidence contours, particularly for the HMF, can potentially be reduced by increasing the bin boundaries and/or by fine-tuning the binning schemes. Our choice of binning is restricted by our simulation resolution. We leave these investigations to future works.

\begin{figure*}
\begin{center}
\includegraphics[width=\textwidth]{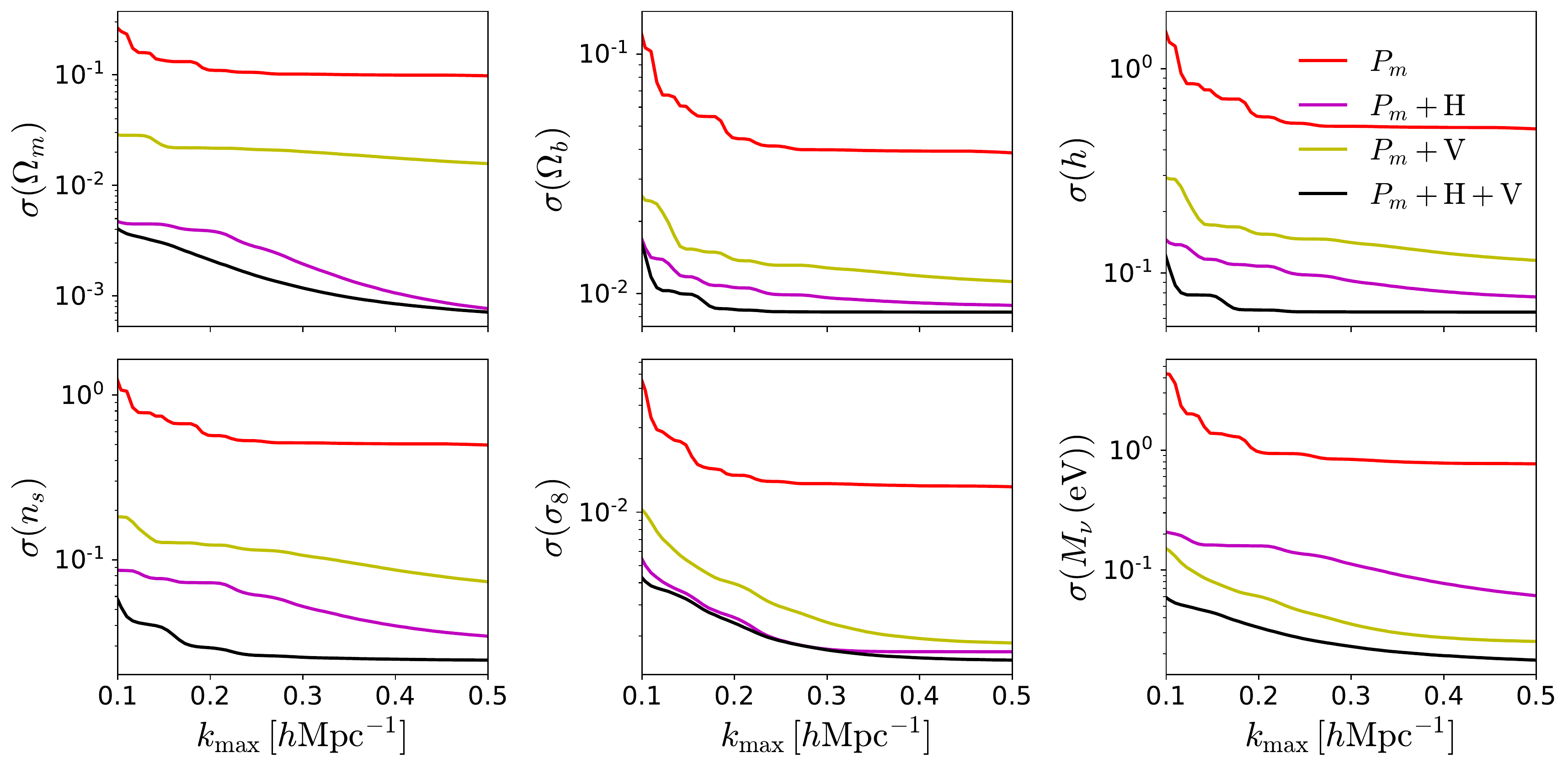}
\caption{\label{fig:kmax}The 1D marginalized error for each of the cosmological parameters as a function of $k_{\rm max}$. We consider 4 scenarios: $P_m$ alone (red), $P_m + {\rm HMF}$ (magenta), $P_m + {\rm VSF}$ (yellow),  and $P_m + {\rm HMF} +  {\rm VSF}$ (black). 
%For $k_{\rm max} = 0.5\,h{\rm Mpc}^{-1}$, the multiplicative reduction in errors achieved by combining all three probes compared to $P_m$ alone are 137, 5, 8, 20, 10, and 43, for $\Omega_m$, $\Omega_b$, $h$, $n_s$, $\sigma_8$, and $M_\nu$, respectively. 
While the constraints from $P_m$ alone saturate at $k_{\rm max} \simeq 0.2\,h{\rm Mpc}^{-1}$, the combined constraints for $M_\nu$ (and $\Omega_m$) continue to improve until $k_{\rm max} = 0.5\,h{\rm Mpc}^{-1}$, and likely beyond.}
\end{center}
\end{figure*}

For a direct comparison to the usual constraints expected from the matter power spectrum, we show the 1D marginalized errors (Eq.~\ref{eq:marg_err}) from different combinations of the probes with $P_m$ in Fig.~\ref{fig:kmax}.  We study how the errors vary with the cutoff scale $k_{\rm max}$.
Combining $P_m$ with either the HMF, VSF, or both, can achieve a significant level of improvement on all 6 parameters. The combination with the HMF is typically more beneficial than the combination with the VSF. The only exception is for $M_\nu$, where the VSF is the better probe to combine with $P_m$. 

While the constraints from $P_m$ alone saturate at around $k_{\rm max} = 0.2\,h{\rm Mpc}^{-1}$ for all parameters, the combined constraints for $M_\nu$ (and $\Omega_m$) continue to improve beyond $k_{\rm max} = 0.5\,h{\rm Mpc}^{-1}$. This can be explained by the breaking of degeneracies when combining probes. It was shown in Fig.~5 of \cite{quijote} that increasing $k_{\rm max}$ beyond  $0.2\,h{\rm Mpc}^{-1}$ leads to a squeezing along the semi-minor axes (i.e,~the most constraining direction) for the $P_m$ ellipses. While this squeezing has little effect on the marginalized error on $M_\nu$ from $P_m$ alone, its effects are manifest when combined with other probes with misaligned contours, resulting in significant tightening of constraints.
Even though the numerical resolution of the \textit{Quijote} simulations prevent us from confidently investigating beyond $k_{\rm max} = 0.5\,h{\rm Mpc}^{-1}$, our results hint that even  tighter constraints could be achieved by including smaller scales.

In Table \ref{tab:errors} we list the errors for $k_{\rm max} = 0.5 h{\rm Mpc}^{-1}$ using different probe combinations. We list the constraints obtained by combining all three probes while (1) only using the diagonals of the covariance matrix (diag), (2) only considering auto-covariance (auto), and (3) considering the full covariance (full). 
We find that using only the diagonal components of the covariance matrix, effectively ignoring both the correlation between the probes and between different bins of the same probe, leads to a factor of 1.7 increase on the error on the neutrino mass. Using only block cross-correlations, i.e. ignoring the correlation between the probes, leads to a factor of 1.2 increase on the error on the neutrino mass. Therefore, to obtain the tightest constraints, it is crucial to model the full covariance matrix. It is interesting to note that when considering the matter power spectrum alone, correlations cause an increase in errors due to the \textit{positive} correlation between different scales (see Fig.~\ref{fig:Covariance}). However, it is the complex correlation structure, notably the anti-correlations, introduced by considering the HMF and VSF that leads to a reduction in error, both for the HMF and VSF individually, and in turn when combining all probes. The association of anti-correlation with the tightening of constraints was also pointed out by \cite{chartier2020carpool}.

In Table \ref{tab:errors} we quantify the improvement of the combined constraints compared to those achieved from $P_m$ alone.
%, given by the ratios of the red and black lines in Fig.~\ref{fig:kmax}. 
We find the improvements to be a factor of 137, 5, 8, 20, 10, and 43, for $\Omega_m$, $\Omega_b$, $h$, $n_s$, $\sigma_8$, and $M_\nu$, respectively. Thus we achieve 43 times tighter constraints on neutrino mass by combining all three probes.
Specifically, the marginalized errors on $M_\nu$ are $0.77 {\rm eV}$ ($P_m$ alone) and $0.018 {\rm eV}$ ($P_m$+HMF+VSF).
%,  while those on $h$ are 0.51 ($P_m$ alone) and 0.064 ($P_m$+HMF+VSF).
We provide an additional plot in Appendix \ref{app:probe_pairs} to show the confidence ellipses when combining only two of the probes at a time.

\begin{table*}
\begin{center} 
 
\resizebox{0.75\textwidth}{!}{\begin{tabular}{l|llllll} \toprule
\multicolumn{7}{c}{Marginalized Fisher Constraints} \\[2pt] \toprule
Probe(s) & $\Omega_m$ & $\Omega_b$ & $h$ & $n_s$ & $\sigma_8$ & $M_\nu ({\rm eV})$ \\ [2pt]\hline\hline
        
$P_m$ & 0.098 & 0.039 & 0.51 & 0.50 & 0.014 & 0.77 \\ 
${\rm HMF}$ & 0.034 & 0.042 & 0.28 & 0.12 & 0.082 & 1.6 \\ 
${\rm VSF}$ & 0.31 & 0.12 & 1.3 & 0.42 & 0.083 & 1.1 \\ \hline
$P_m+{\rm HMF}$ & 0.00077 & 0.0089 & 0.076 & 0.034 & 0.0016 & 0.061 \\ 
$P_m+{\rm VSF}$ & 0.016 & 0.011 & 0.12 & 0.074 & 0.0018 & 0.025 \\ 
${\rm HMF}+{\rm VSF}$ & 0.0063 & 0.037 & 0.23 & 0.10 & 0.0069 & 0.096 \\ \hline
$P_m+{\rm HMF}+{\rm VSF}$ (diag) & 0.0015 & 0.0088 & 0.066 & 0.028 & 0.00061 & 0.031 \\ 
$P_m+{\rm HMF}+{\rm VSF}$ (auto) & 0.0015 & 0.0086 & 0.071 & 0.033 & 0.0016 & 0.025 \\\hline
$P_m+{\rm HMF}+{\rm VSF}$ (full) & 0.00071 & 0.0084 & 0.064 & 0.025 & 0.0015 & \textbf{0.018} \\ \hline\hline
Multiplicative improvement & 137 & 5 & 8 & 20 & 10 & 43 \\ \hline
        
%$P_m$ & 0.098 & 0.039 & 0.507 & 0.497 & 0.014 & 0.768 \\ 
%${\rm HMF}$ & 0.034 & 0.042 & 0.279 & 0.117 & 0.082 & 1.601 \\ 
%${\rm VSF}$ & 0.314 & 0.117 & 1.265 & 0.418 & 0.083 & 1.145 \\ \hline
%$P_m+{\rm HMF}$ & 0.00077 & 0.009 & 0.076 & 0.034 & 0.002 & 0.061 \\ 
%$P_m+{\rm VSF}$ & 0.016 & 0.011 & 0.115 & 0.074 & 0.002 & 0.025 \\ 
%${\rm HMF}+{\rm VSF}$ & 0.0063 & 0.037 & 0.225 & 0.103 & 0.0069 & 0.096 \\ \hline
%$P_m+{\rm HMF}+{\rm VSF}$ (diag) & 0.0015 & 0.009 & 0.066 & 0.028 & 0.00061 & 0.031 \\ 
%$P_m+{\rm HMF}+{\rm VSF}$ (auto) & 0.0015 & 0.009 & 0.071 & 0.033 & 0.0016 & 0.025 \\ \hline
%$P_m+{\rm HMF}+{\rm VSF}$ (full) & 0.00071 & 0.0084 & 0.064 & 0.025 & 0.0015 & \textbf{0.018} \\ \hline
\end{tabular}} 
\caption{\label{tab:errors} Marginalized errors of cosmological parameters for $k_{\rm max} = 0.5 h{\rm Mpc}^{-1}$ using different probe combinations. Note, we list the constraints obtained by combining all 3 probes while: 1) only using the diagonals of the covariance matrix (diag), 2) only considering auto-covariance (auto), and 3) considering the full covariance (full). We highlight in bold the full constraints on the sum of the neutrino masses. We also list the multiplicative improvement in the constraints from the full covariance compared to those from $P_m$ alone.} 
\end{center}
\end{table*}

%In order to verify the robustness of our results we have carried out some tests which are documented in Appendix \ref{app:robustness}. Specifically we consider the effects of using fewer simulations to compute the covariance matrix and derivatives, as well as changing the binning and dynamical range of the independent variables.

%%%%%%%%%%%%%%%%%%%%%%%%%%%%%%%%%%%%%%%%%%%%
\section{Discussion and Conclusions}
\label{sec:Conclusions}

Upcoming galaxy surveys will map large volumes of the Universe at low redshifts, with the potential to drastically improve our understanding of the underlying cosmological model. With the unprecedentedly precision achievable by these surveys, it is expected that a very large amount of cosmological (and astrophysical) information will lie in the mildly to fully non-linear regime, where analytic methods are often intractable. It remains an open question which observable(s) will lead to the tightest bounds on the cosmological parameters. 

In this paper, we use the \textit{Quijote} simulations, based on the Fisher formalism, to
quantify the information content embedded in the non-linear matter power spectrum, the halo mass function, and the void size function, both individually and when combined, at $z=0$. We find that the HMF and VSF
%, individually, do not give tighter constraints than the matter power spectrum. However, they 
have different degeneracies to each other and to the matter power spectrum, particularly in the $M_\nu$--$\sigma_8$ plane (Figs.~\ref{fig:Fisher_individual} \& \ref{fig:Mnu-s8}). In terms of measuring neutrino mass, we find the void size function to be the more complementary probe to combine with the matter power spectrum. This is consistent with findings that void properties are particularly sensitive to matter components that are less clustered, such as neutrinos \citep{Massara2015,Kreisch2019}.

By combining the non-linear matter power spectrum ($k_{\rm max}=0.5~h{\rm Mpc}^{-1}$), with the halo mass function ($M\gtrsim2\times10^{13}~h^{-1}M_\odot$), and the void size function ($R\geqslant10.4 h^{-1}{\rm Mpc}$), we achieve significantly tighter constraints on the cosmological parameters compared to $P_m$ alone (Fig.~\ref{fig:kmax}). In particular, we find that with a volume of just $1~(h^{-1}{\rm Gpc})^3$, the error on the sum of neutrino masses from the combined probes is at the $0.018 {\rm eV}$ level, compared to $0.77 {\rm eV}$ from the matter power spectrum alone --- a factor of 43 improvement. 
We emphasize that this value mainly demonstrates the information content in the late-time statistics, and they are not forecasts for any particular survey. 

Also of particular interest is the factor 137 improvement in the error on $\Omega_m$. This is driven by the information in the HMF, and gives a marginalized error of $\sigma(\Omega_m)=7.1\times10^{-4}$, which is almost 100 times smaller than the error obtained from a joint large-scale structure analysis by DES Y1 \citep[$\sigma(\Omega_m)\approx 0.04$,][]{DES_6x2}, and 8 times smaller than Planck 2018,  \citep[$\sigma(\Omega_m)\approx 5.6\times10^{-3}$ (TT,TE,EE+lowE+lensing+BAO),][]{planck2018}. 
In addition, we found $\sigma(h) = 0.064$ by combining the three probes, which is 8 times tighter than the constraints from the matter power spectrum alone. This could provide a new angle to investigate the Hubble tension. 

There are several caveats in this work. Firstly, we assumed perfect knowledge of the three-dimensional spatial distribution of the underlying matter field in real-space. However, in reality, one observes either tracers of  the matter field in redshift-space, or the projected matter field through lensing. Therefore, additional links must be made to bridge the galaxy--matter connection and the 2D lensing--3D matter distribution gaps. 
This effect is also relevant for voids:
in this work we considered voids in the 3D matter field, which is not something current surveys are able to observe directly. Detecting voids in the matter field from photometric (2D lensing) data has been considered in works such as \citep{Pollina_2019, Davies_2020}.
Alternatively, one can measure voids in the 3D halo field \citep[see, e.g.][]{Nadathur_2016, Contarini_2019}. 
If we were to instead have considered voids in the 3D CDM field, the combined error on $M_\nu$ slightly degrades to 0.025eV. However, considering voids in the CDM field versus halo field can lead to non-trivial differences in void properties, which might increase or decrease constraints \citep{Kreisch2019}. We will consider voids in the halo field in a future work.

A further note regarding voids is that there are various conventions when it comes to defining voids \citep[see, e.g.][]{Platen_2007, VIDE}. It would thus be interesting further work to consider how the choice of void finder impacts constraints. A different void finder may be able to extract additional information compared to the spherical void finder applied here.

Another limitation of this work is that our simulations consider only gravitational interactions and hence ignore baryonic effects which can impact the small-scale matter distribution. This is particularly relevant for both clustering and halos \citep[see, e.g.][and references therein]{CAMELS, cromer20211,Debackere_2021}, while it is expected that baryons have a lower impact on voids \citep{Paillas2017}. Furthermore, halo clustering is influenced by various properties, such as spin, concentration, and velocity anisotropy, which have not been considered in this work \citep[see, e.g.][]{Wechsler_2006, Gao_2007, Faltenbacher_2009, Lacerna_2012, Lacerna_2014, Paranjape_2018, Shi_2018}.  

Additionally, we have neglected super-sample covariance \citep{Takada_2013,Li_2014}, which could modify the errors reported in this work.

%Another thing to consider is the $1 h^{-1}{\rm Gpc}$ box size used in this work. Unlike halos, voids are extended objects and may exceed even the size of this box when using non-spherical void finders.

We also note that the constraints obtained here may be overly conservative due to the limited number and resolution of simulations available. Firstly, this means that the number of bins used are likely suboptimal. Second, applying more aggressive bounds on the observables, e.g.~a higher $k_{\rm max}$, a larger halo mass range, or a larger void size range, would likely also reduce the combined constraints. 
%We also note that redshift space distortions (RSD) are known to increase the amount of information on large scales -- while it is not known that this carries over to small scales, we expect that including RSD would decrease our quoted errors. 
Third, we only considered a single redshift, $z=0$: in practice, surveys measure $z>0$ where the universe is more linear and the constraints will thus be weaker, however, combining multiple redshifts could tighten the constraints as found in works such as \cite{liu&madhavacheril2019}.
Fourth, we considered a volume of only $1\, (h^{-1}{\rm Gpc})^3$, whereas surveys such as Euclid and DESI will cover volumes of around $10^2\, (h^{-1}{\rm Gpc})^3$, so, conservatively, the error on the parameters will shrink by a factor of $1/\sqrt{10^2}=0.1$.
Fifth, we have only considered three probes; using the same observations, one can derive other statistics such as the bispectrum, void profile, and BAO, which could be combined with the statistics considered here to further break degeneracies. Finally, considering redshift space distortions would also tighten constraints as neutrinos are distinguishable from CDM via their higher thermal velocity.

%To give a quantitative idea of what this means: if bias,  baryonic effects, redshift-space distortions,binning, etc., combine to give an overall increase in the errors of less than 300\%, one would obtain a $< 0.1 \times 4 \times 0.031 {\rm eV} \approx 0.012 {\rm eV}$ constraint on the minimum sum of the neutrino masses. Given the theoretical lower bound is $\sim 0.06 {\rm eV}$ this would result in a $\gtrsim 5 \sigma$ detection. A similar scaling for the Hubble paramter would give an error of $\sigma(h) < 0.1 \times 4 \times 0.064 = 0.022$, which of order a few percent ($0.026/0.7 \simeq 3.6 \%$).

%Hence, if less than 90\% of the available information is erased after factoring in bias, redshift-space distortions, baryonic effects, binning, etc., one would obtain a $< 0.031 {\rm eV}$ constraint on the minimum sum of the neutrino masses. Given the theoretical lower bound is $\sim 0.06 {\rm eV}$ this would results in a $< 2 \sigma$ detection.

We have demonstrated that combining multiple probes of cosmological structure using their full covariance matrix provides remarkably tight constraints on the cosmological parameters, and helps extract much additional information from small scales. In particular, we have shown that there is, in principle, sufficient information to measure the sum of the neutrino masses at the minimum mass of $0.06 \,{\rm eV}$. Our results are in good agreement with \cite{Sahl_n_2019} who found that combining halo and void abundances can yield $\mathcal{O}(0.01 \,{\rm eV})$ constraints on the neutrino mass. This approach opens a promising pathway to measure neutrino mass, potentially without relying on CMB-based measurements which require accurate knowledge of the optical depth, $\tau$. In addition, comparing constraints from different combinations of observables, e.g., CMB+$P_m$ and $P_m$+HMF+VSF, will help identify systematic issues and provide robust evidence for any discovery. We thus hope our work will motivate galaxy survey collaborations to build the simulations and analytic tools necessary to implement this approach on upcoming observational data.

\section*{ACKNOWLEDGEMENTS} 
We thank Ravi Sheth for useful conversations on the early stages of this project, and Alice Pisani for fruitful discussion regarding voids.
The \textit{Quijote} simulations can be found at \url{https://github.com/franciscovillaescusa/Quijote-simulations}. The analysis of the simulations has made use of the \textit{Pylians} libraries, publicly available at \url{https://github.com/franciscovillaescusa/Pylians3}. The work of D.N.S., B.D.W., and S.H. is supported by the Simons Foundation. L.V. acknowledges support from the European Union Horizon 2020 research and innovation program ERC (BePreSySe, grant agreement 725327) and MDM-2014-0369 of ICCUB (Unidad de Excelencia Maria de Maeztu). M.V. is supported by INFN PD51 Indark grant and by the agreement ASI-INAF n.2017-14-H.0.

%%%%%%%%%%%%%%%%%%%%%%%%%%%%%%%%%%%%%%%%%%%%
\begin{appendix}

\section{A. Robustness of results to numerical systematics}
\label{app:robustness}

\begin{figure*}
\begin{center}
\includegraphics[width=0.99\textwidth]{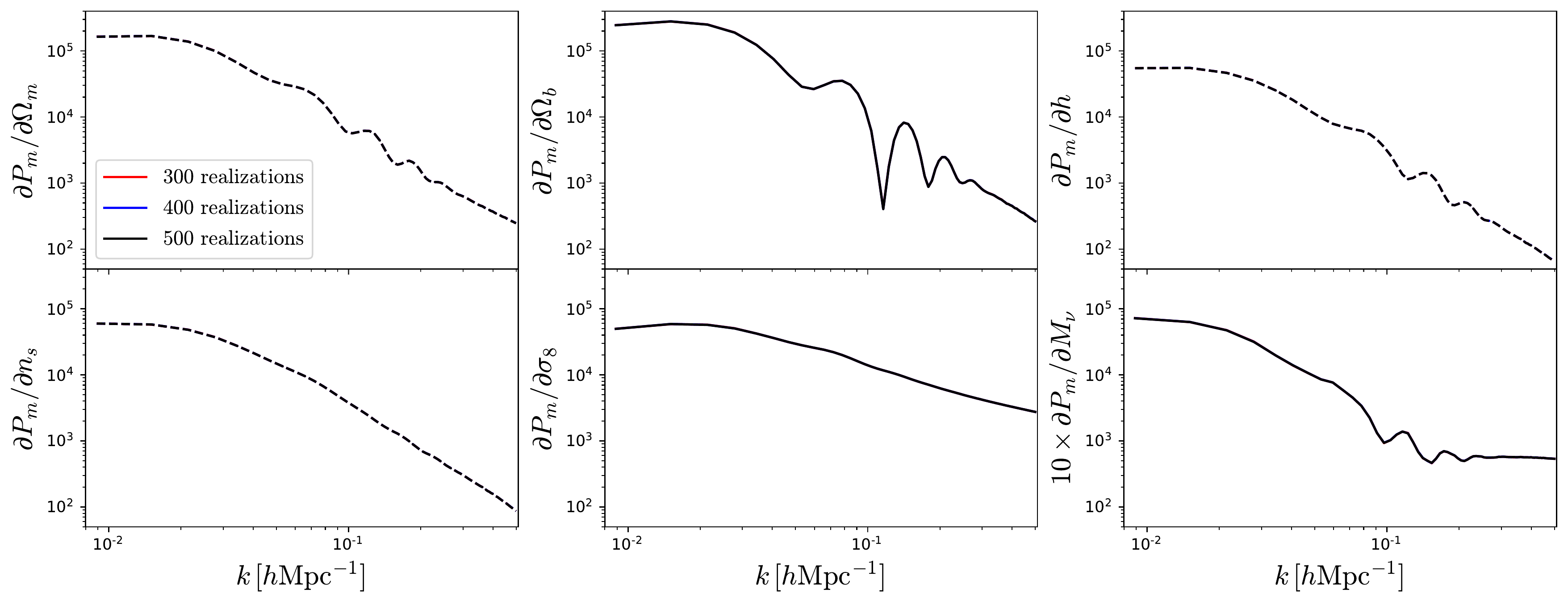}\\
\vspace{0.5cm}
\includegraphics[width=0.99\textwidth]{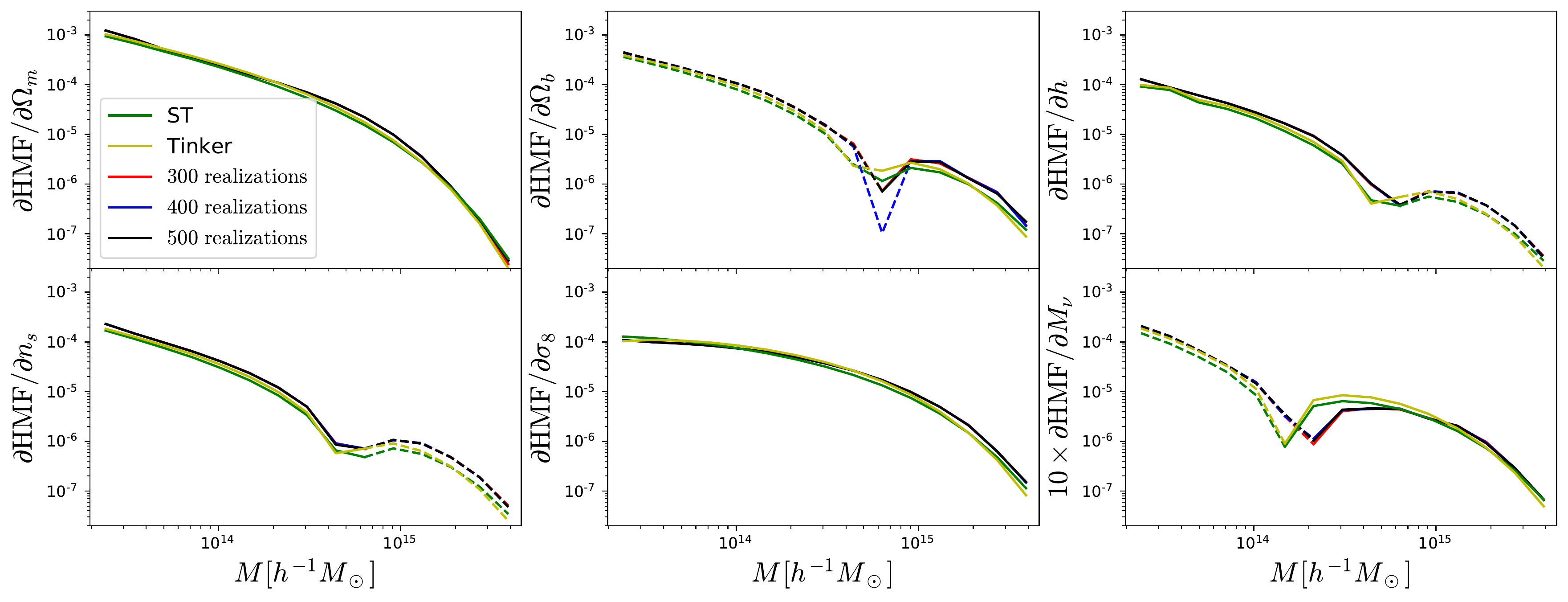}\\
\vspace{0.5cm}
\includegraphics[width=0.99\textwidth]{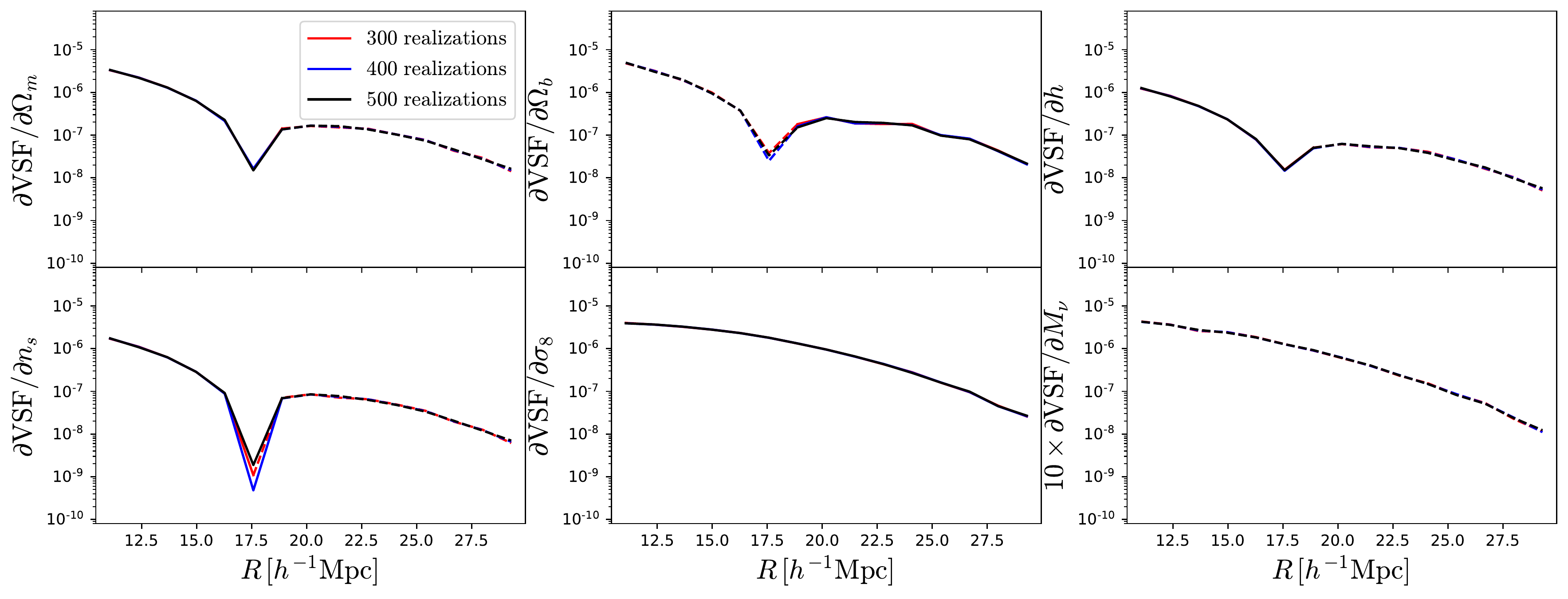}
\caption{Derivatives of the matter power spectrum (top), halo mass function (middle), and void size function (bottom) with respect to the different cosmological parameters at $z=0$. We show results when the mean values are estimated using 300 (red), 400 (blue), and 500 realizations (black). Solid/dashed lines indicate that the value of the derivative is positive/negative. While the derivatives for the matter power spectrum are well converged already with 300 realizations, more simulations are required for halos and voids.}
\label{fig:derivatives}
\end{center}
\end{figure*}

\begin{figure*}
\begin{center}
\includegraphics[width=0.49\textwidth]{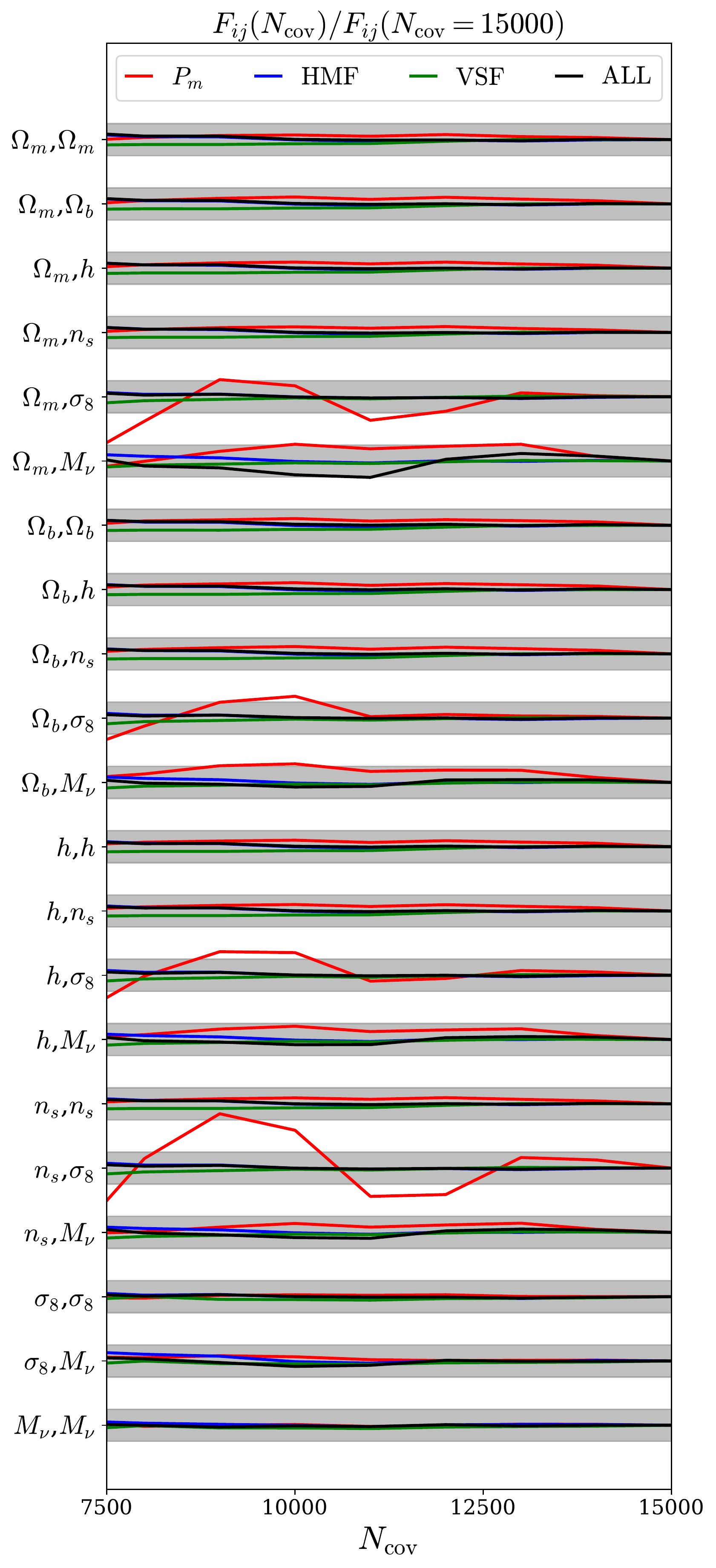} \includegraphics[width=0.49\textwidth]{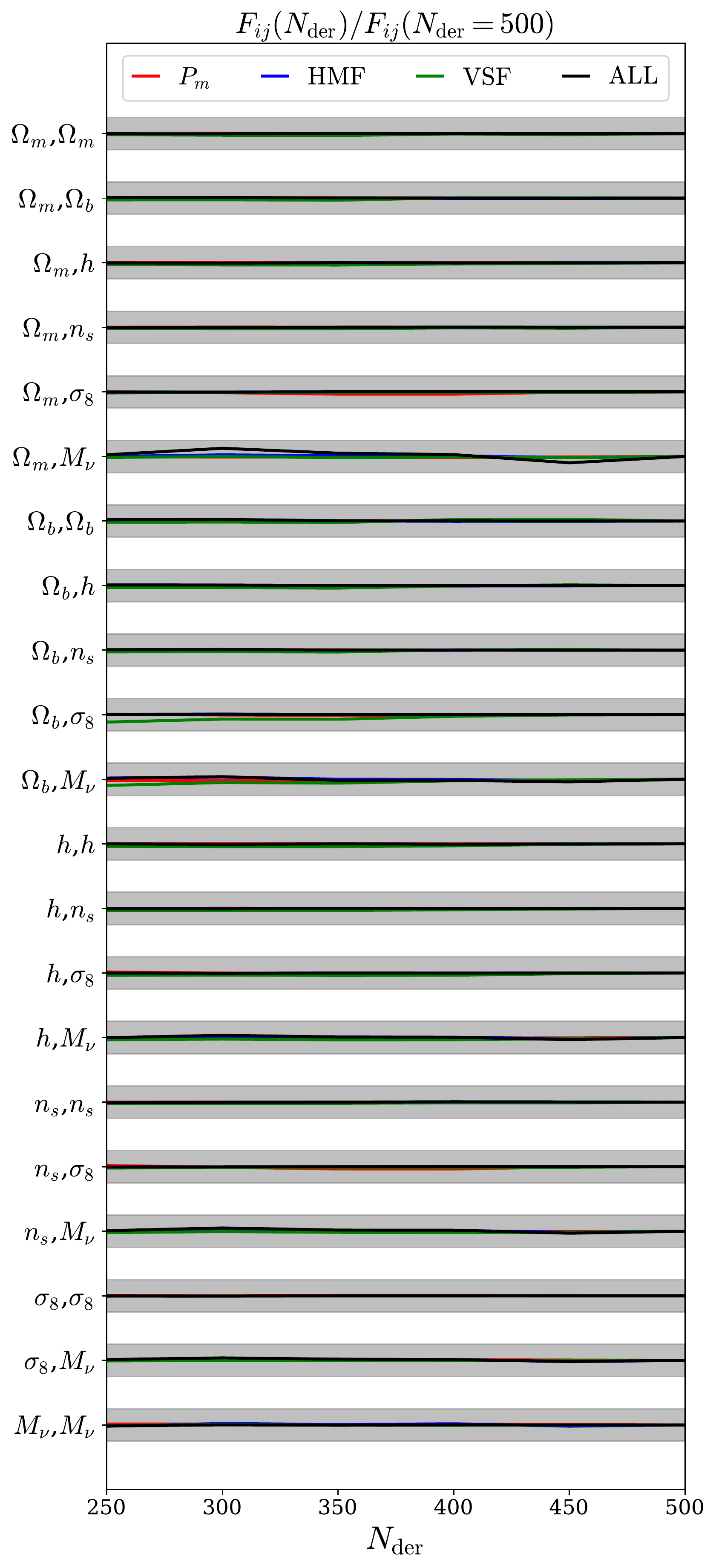}
\caption{{Left:} Convergence of all Fisher matrix components as a function of number of simulations used to compute the covariance matrix, $N_{\rm cov}$. Each line shows the ratio between the Fisher matrix elements computed using $N_{\rm cov}$ simulations and 15,000 simulations (as used in the paper).
{Right:} Convergence of all Fisher matrix components as a function of number of simulations used to compute derivatives, $N_{\rm der}$. Each line shows the ratio between the Fisher matrix elements computed using $N_{\rm der}$ simulations and 500 simulations for each cosmology (as used in the paper).
In both cases, we plot the Fisher matrix components for $P_m$ (red), the HMF (blue), the VSF (green), and the combined probes (black). The gray bands correspond to the $\pm 5 \%$ interval. While there is some noise in the $\sigma_8$ component of the Fisher matrix for $P_m$ as a function of $N_{\rm cov}$, good convergence is achieved by 15,000. Likewise the Fisher matrix is well converged as a function of $N_{\rm der}$. Crucially, the Fisher matrix elements for the combined probes (black) all show good convergence.
}
\label{fig:Fcoldev}
\end{center}
\end{figure*}

In this section we verify the stability of our results to reduction in the number of simulations used to compute the covariance matrix and derivatives.
%, as well as changing the binning.
% and dynamical range of the independent variables. 
In Fig.~\ref{fig:derivatives} we show the derivatives of the matter power spectrum (top), halo mass function (middle), and void size function (bottom) with respect to the cosmological parameters when using a different number of realizations. For the matter power spectrum, the derivatives are already converged when the mean values for each model are computed with 300 realizations. Results are slightly noisier for the halo mass function and the void size function, but still sufficiently converged by 500 realizations.

Next, we comment on the convergence of our simulated results with theory.
The convergence of the matter power spectrum in \textit{Quijote} has been thoroughly tested \citep{quijote, Aviles_2020,Hahn_2020}. For the void size function there is no theoretical formula accurate enough to compute derivatives, but we have checked results are robust to the parameters used in the void finder. Therefore, we only compare our measured HMF to theoretical predictions. For the HMF, we plot the theoretical predictions of Sheth-Tormen (ST) \citep{Sheth_1999, Sheth-Tormen} and Tinker \citep{Tinker_2008}. We use the prescription of \cite{Costanzi_2013} in the case of massive neutrino cosmologies by replacing $\Omega_m \rightarrow \Omega_{cb}$ and $P_m \rightarrow P_{cb}$ as neutrinos have negligible contribution to halo mass. There is good agreements between these predictions and \textit{Quijote}. We have also checked that there is good agreement for different choice of step size (not shown). Note that these theoretical formulae provide a guideline rather than exact predictions, as they were fitted to simpler simulations or calibrated on spherical overdensity halos, as opposed to FoF here.
%, as neither posses the required accuracy to compute derivatives for the Fisher analysis: Sheth-Tormen is somewhat outdated, while Tinker is a fitting formula for spherical overdensity (SO) halos, as opposed to FoF here. It is interesting to note that the Quijote derivatives are, on average, slightly larger in magnitude than theory; we find this leads to $\sim 10\%$ tighter marginalized errors compared to when using the theoretical derivatives directly in the Fisher analysis.

In Fig.~\ref{fig:Fcoldev} we show the convergence of the Fisher matrix elements with respect to the number of realizations used to compute the covariance, $N_{\rm cov}$, and derivatives $N_{\rm der}$. We consider the Fisher matrix components for $P_m$ (red), the HMF (blue), the VSF (green), and the combined probes (black). The gray bands corresponds to the $\pm 5 \%$ interval. While there is some noise in the $\sigma_8$ component of the Fisher matrix for $P_m$ as function of $N_{\rm cov}$, good convergence is achieved by 15000. Likewise the Fisher matrix is well converged as a function of $N_{\rm der}$. Crucially, the Fisher matrix elements for the combined probes (black) all show good convergence. Note that when combining probes we scale the power spectrum by a factor of $10^{-10}$ to ensure that the condition number of the covariance matrix is sufficiently low for accurate inversion.

Finally, we comment on the choice of finite difference scheme used to compute the derivative of probes with respect to $M_\nu$. Throughout the paper we used Eq.~\ref{eq:error_nu2} with $\delta M_\nu = 0.2 \, {\rm eV}$, thus making use of simulations with $M_\nu = 0, 0.2$, and 0.4 eV. Using this scheme we found the full combined constraint on $M_\nu$ is 0.018 eV, as shown in Table \ref{tab:errors}. To illustrate robustness to this choice of finite difference scheme, we also performed the analysis using Eq.~\ref{eq:error_nu2} with $\delta M_\nu = 0.1 \, {\rm eV}$ and found it to give an identical constraint of 0.018 eV. Additionally, we tried a forward difference scheme between $M_\nu = 0$ and 0.1 eV, which also gave identical constraints. Hence, the results are consistent with the choice of finite difference scheme. We do also note that since the joint constraints on the parameters given in Table \ref{tab:errors} are smaller than the step sizes used to compute derivatives, it would be interesting to investigate the effect of smaller step sizes on the joint constraints. This would reduce the error in the numerical derivatives, and thus may slightly modify the joint constraints.

Given these results, we believe that our conclusions are robust against potential numerical systematics. We note again that our bin configuration has been chosen with these results in mind, to ensure sufficiently converged derivatives and Fisher matrix components, but in principle one could consider more bins over a wider range to potentially obtain tighter constraints.

%%%%%%%%%%%%%%%%%%%%%%%%%%%%%%

\vspace{-.075cm}
\section{B. Combining two probes at a time}
\label{app:probe_pairs}

\begin{figure*}
\begin{center}
\includegraphics[width=\textwidth]{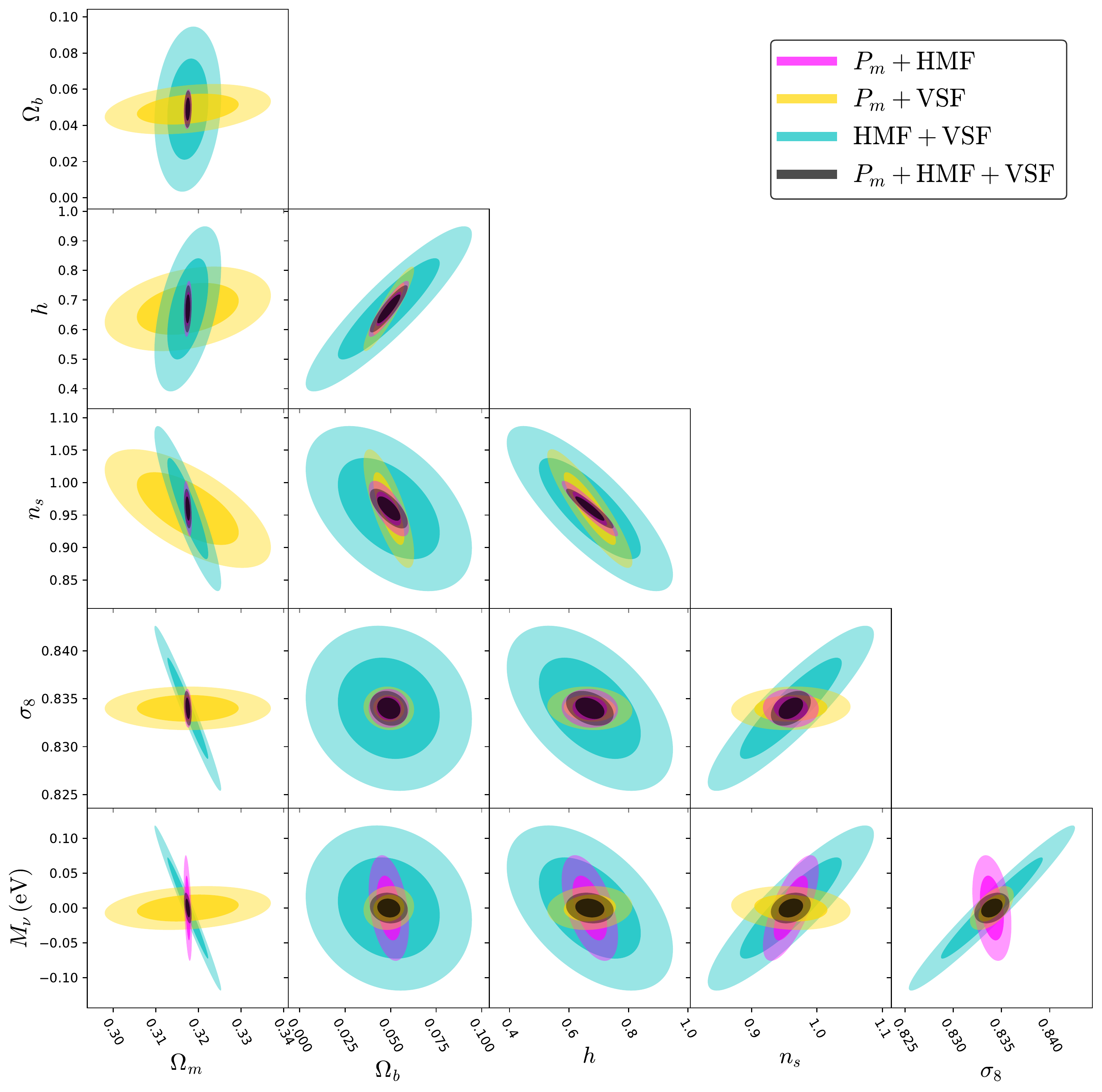}
\caption{Same as Fig.~\ref{fig:Fisher_individual} but for pair combinations of the probes: power spectrum + halo mass function (magenta), power spectrum + voids size function (yellow), halo mass function + void size function (cyan) and power spectrum + halo mass function + void size function (black).}
\label{fig:Fisher_combined}
\end{center}
\end{figure*}

Fig.~\ref{fig:Fisher_combined} shows the 2D Fisher contours to illustrate the effects of only combining two out of the three probes at a time.
In most cases, the constraints obtained by combining the halo mass function with the void size function are the weakest, indicating that it is important to use the information from the non-linear matter power spectrum to break degeneracies.\\

%In general, the combination of the power spectrum with the halo mass function provides similar or tighter constraints than the power spectrum plus the void size function, with the exception of $n_s$.
%For neutrinos, we can see that the tight constraints can only be achieved by using the three different statistics. The combination of two probes that best constraint their masses is the non-linear matter power spectrum plus the halo mass function.
%

\end{appendix}

%\newpage

\bibliography{references}{}
\bibliographystyle{hapj}

\end{document}